\documentclass{article}
\usepackage{amsmath}
\usepackage{amssymb}
\usepackage{amsfonts}
\usepackage{epic}
\usepackage{eepic}
\usepackage{epsfig}
\usepackage{graphicx,caption}
\usepackage{accents}

\def\tr{{\mathrm{Tr}}}
\def\det{{\mathrm{Det}}}
\def\cL{{\mathcal L}}
\def\cM{{\mathcal M}}

\def\N{{\mathbb N}}
\def\Z{{\mathbb Z}}
\def\abs#1{{\mid #1 \mid}}
\def\max{{\text{max}}}

\def\q{{r}}
\def\s{{\bf s}}
\def\a{{\bf a}}
\def\b{{\bf b}}
\def\u{{\bf u}}
\def\map{\phi}
\def\mapt{\varphi}
\def\Pm{\sigma}
\def\Rm{\tau}
\def\t#1{\widetilde{#1}}
\def\h#1{\widehat{#1}}
\def\ut#1{\underset{\widetilde{}}{#1}}
\def\uh#1{\underset{\widehat{}}{#1}}

\newcounter{NN}
\newtheorem{proposition}[NN]{Proposition}
\newtheorem{theorem}[NN]{Theorem}

\newtheorem{lemma}[NN]{Lemma}

\def\proof{\noindent {\bf Proof:} }
\def\qed{{\hfill $\square$\\ \noindent}}

\begin{document}
\bibliographystyle{plain}
\title{The staircase method: integrals for periodic reductions of integrable lattice equations}
\author{Peter H.~van der Kamp, G.R.W. Quispel}
\date{email: P.vanderKamp@LaTrobe.edu.au\\[3mm]
Department of Mathematics and Statistics\\
La Trobe University \\
Victoria 3086, Australia\\[5mm]
date: \today \\[5mm]
 }
\pagestyle{plain} \maketitle

\begin{abstract}
We show, in full generality, that the staircase method \cite{PNC,QCPN} provides integrals
for mappings, and correspondences, obtained as traveling wave reductions of (systems of)
integrable partial difference equations. We apply the staircase method
to a variety of equations, including the Korteweg-De Vries equation, the five-point Bruschi-Calogero-Droghei
equation, the QD-algorithm, and the Boussinesq system. We show that, in all these cases,
if the staircase method provides $r$ integrals for an $n$-dimensional mapping, with $2r<n$,
then one can introduce $q\leq 2r$ variables, which reduce the dimension of
the mapping from $n$ to $q$. These dimension-reducing variables are obtained as
joint invariants of $k$-symmetries of the mappings. Our results support the idea that
often the staircase method provides sufficiently many integrals for the periodic reductions
of integrable lattice equations to be completely integrable. We also study reductions
on other quad-graphs than the regular $\Z^2$ lattice, and we prove linear
growth of the multi-valuedness of iterates of high-dimensional correspondences
obtained as reductions of the QD-algorithm.
\end{abstract}

\section{Introduction}
The field of integrable partial difference equations emerged
in the late nineteen seventies, early eighties \cite{AL,DJM,H,NQC,QNCL}.
An important, and well-studied, class of partial difference equations is the
class of (scalar) equations that are defined on the elementary squares of a
lattice. An example of an integrable equation in this class is the lattice potential
Korteweg-de Vries (pKdV) equation
\begin{equation} \label{kdv}
(u_{l,m}-u_{l+1,m+1})(u_{l+1,m}-u_{l,m+1})=\alpha,
\end{equation}
in which the linear terms are transformed away. Such equations are part of a
multi-dimensional family of mutually consistent partial difference equations
\cite{NC,NW}. A classification with respect to multi-dimensional consistency
has been achieved recently \cite{ABS1,ABS2}.

For lattice equations on the square, an initial value
problem can be posed on a so called {\em staircase}: a connected
path which is nondecreasing, or nonincreasing. In \cite{PNC} initial values are given
at lattice points $u_{l,l}$ and $u_{l+1,l}$ which satisfy the periodicity $u_{l,m}=u_{l+p,m+p}$.
By doing so, the partial difference equation ($P\Delta E$) reduces to a multi-dimensional mapping.
The authors of \cite{PNC} used the linear spectral problem (Lax pair) of the lattice pKdV
equation to derive a set of polynomial invariants for this mapping. They
constructed a so called monodromy matrix, which is an ordered product of Lax matrices along
the staircase over a one-period distance. This method, nowadays known as
{\em the staircase method}, is an important tool in proving complete integrability
in the sense of Liouville-Arnold. Here, a $2n$-dimensional mapping is said to be
completely integrable if it admits $n$ functionally independent integrals
in involution with respect to a symplectic form \cite{BRSZ,Ves}. Thus, the number of
integrals should be at least equal to half the dimension of the mapping.

In \cite{CNP,NPC} the authors established the involutivity of the
integrals for the mappings they introduced in \cite{PNC}. Similar
results have been obtained for mappings derived from the lattice
Gel'fand-Dikii hierarchy \cite{NPCQ} and for reductions of the
time-discrete versions of the Bogoyavlensky equations \cite{PN}. In \cite{QCPN} more general
staircases were given, corresponding to the {\em $\s$-periodicity
condition:}\footnote{Our notation differs from the one used in \cite{QCPN},
where $s_1=z_2$ and $s_2=-z_1$.}
\begin{equation} \label{pc}
u_{l,m}=u_{l+s_1,m+s_2},
\end{equation}
where $s_1,-s_2\in\N^+$ are relatively prime integers. The authors
also suggested considering general $s_1,-s_2\in\N^+$, see the third
concluding remark in that paper. In recent work \cite{K,RKQ} we have
provided a unified picture for $\s$-periodic reductions, with
nonzero $\s=(s_1,s_2)\in\Z \times \Z$. In \cite{RKQ} it was shown
how, under periodicity condition (\ref{pc}), any lattice equation
$f(u_{l,m},\cdots)=0$ reduces to a system of $\q$ ordinary
difference equations $f(v_n^p,\cdots)=0$, $p=0,\ldots,\q-1$, where
$\q$ is the greatest common divisor of $s_1$ and $s_2$. Also it was
proved that the monodromy matrix, denoted $\cL$, is one of the Lax
matrices for the reduction, that is, there exists a matrix $\cM$
such that for (periodic) solutions of the system the following
holds,
\begin{equation} \label{ola}
\cL_n\cM_n = \cM_n \cL_{n+1}.
\end{equation}

In \cite{K} a geometric description of $\s$-reduction has been given. It was shown that
for all $\s$ there exists a well-posed, or nearly well-posed, $\s$-periodic initial value problem,
for any given scalar lattice equation on some arbitrary stencil of lattice points. We expect
something similar to hold for systems of lattice equations, cf. \cite{SNK}. Combining the two results;
given the existence of a (nontrivial) periodic solution, after multiplying equation (\ref{ola}) by $\cM_n^{-1}$,
we may conclude that the trace of $\cL$ is an invariant of the mapping $n\mapsto n+1$.
In section \ref{sm} we provide a direct proof, in the spirit of
the original work \cite{QCPN}, that the staircase method applies to any given system of lattice
equations, if a Lax-pair and a (nearly) well-posed periodic initial value problem are known.

Note, for equations on a square the monodromy matrix is defined on the staircase, and the initial
conditions are given at all points of the same staircase. However, for equations on other stencils,
and for systems of equations, the monodromy matrix is still given on the staircase, but the initial
conditions no longer correspond to the points on the staircase.

The trace of the monodromy matrix $\cL$ depends on a spectral parameter, arising from the Lax representation
of the P$\Delta$E. By expanding in this parameter we obtain a number of integrals. In relation to establishing
the complete integrability of a mapping (or correspondence) obtained by periodic reduction a first
question to ask is: {\em does the staircase method yield sufficiently many functionally independent
integrals?}

For the reductions we perform in section \ref{sbcd}, of the Bruschi-Calogero-Droghei
equation \cite{BCD}
\[
(u_{l,m}-u_{l,m-1})(u_{l,m}-u_{l-1,m-1})=(u_{l,m}-u_{l,m+1})(u_{l,m}-u_{l+1,m+1}),
\]
the number of integrals is exactly half the dimension of the mapping. For the one-parameter
families of reductions we perform in section \ref{sqd}, of the QD-system \cite{PGR}
\[
e_{l,m+1} + q_{l+1,m+1} = q_{l+1,m} + e_{l+1,m}, \quad e_{l,m+1} q_{l,m+1} = q_{l+1,m}e_{l,m},
\]
we have verified that the number of integrals is one more than half the dimension of the mapping.
In other cases, there are fewer integrals than half the dimension of the mapping obtained
by periodic reduction. For example, performing periodic reductions of the lattice pKdV equation
(\ref{kdv}) we find ($2n+1$)-dimensional mappings and ($2n+2$)-dimensional mappings, for which
the staircase method provides only $n$ integrals, see section \ref{skdv}. It turns out that when
the dimension of the mapping is $2n+1$ it can be dimensionally reduced by 1, whereas when the dimension is
$2n+2$ it can be reduced by 3. All $n$ integrals survive the dimensional reduction and we can conclude that the
dimensionally reduced mappings posses sufficiently many integrals for complete integrability.

To distinguish the two kinds of reductions we say {\em $\s$-reduction} for a periodic reduction of a
lattice equation with period $\s\in\Z \times \Z$ to a multi-dimensional mapping, and we say {\em $d$-reduction}
for a reduction of order which reduces the dimension of a mapping by $d\in\N$.

In section \ref{sB} we show how to pose $\s$-periodic initial value problems for the Boussinesq system
\cite{NPCQ,TN}
\begin{align*}
w_{l+1,m} + v_{l,m}   &= u_{l,m} u_{l+1,m},\\
w_{l,m+1} + v_{l,m}   &= u_{l,m} u_{l,m+1}, \\
w_{l,m} + v_{l+1,m+1} &= u_{l,m} u_{l+1,m+1} + \frac{\gamma}{u_{l+1,m} - u_{l,m+1}}.
\end{align*}
Performing $\s$-reduction with $\s=(n-1,1)$ we get a $2n$ dimensional mapping. For these mappings
we verified, for all $n\leq 17$, that the staircase method provides $n-1$ integrals unless $3$ divides
$n$ in which case it provides only $n-3$ functionally independent integrals. We show, for all $n$,
that the mapping can be $6$-reduced if $3$ divides $n$, and that the mapping can be
$2$-reduced otherwise.

These examples suggest that if the staircase method provides $r$ integrals for an $n$-dimensional
mapping, with $2r<n$, then the mapping can be $d$-reduced, with $d\geq n-2r$. However, we do not
claim the above statement is true in general; in examples given in \cite{SNK} the staircase method
gives integrals of the form $JJ^\prime$ where $J$ is a 2-integral, and it does not produce the integral $J+J^\prime$.
Recall, a function $J$ is an $k$-integral, or $k$-symmetry, of a mapping if it is an integral, or symmetry, of the
$k^\text{th}$ power of that mapping \cite{HBQC}. If one has one $k$-integral, then one can construct $k$ of them, or,
even better, $k$ integrals. For example, it is easy to see that $J^{\prime\prime}=J$ implies that both
$JJ^\prime$ and $J+J^\prime$ are integrals.
In all cases considered in this paper, the $d$-reduction is performed by introducing $n-d$ new variables,
which can be obtained as the joint invariants of symmetries, or $k$-symmetries, of the mapping,
which in turn are obtained from point-symmetries of the partial difference equation. This will be
explained in section \ref{dr}.

Recently, in \cite{AV}, a geometric criterion was given for the well-posedness of initial
value problems on quad-graphs. In section 5 we will show that for `regular' quad-graphs,
those that permit periodic solutions, the staircase method can be applied. We study reductions
of equation $H3_{\delta=0}$ from \cite{ABS1}, which on a $\Z^2$-lattice would look like
\[
p(u_{l,m}u_{l+1,m}+u_{l,m+1}u_{l+1,m+1})=q(u_{l,m}u_{l,m+1}+u_{l+1,m}u_{l+1,m+1}).
\]
We will consider two different quad-graphs, namely Figure 9d and 9e in \cite{AV}. These
quad-graphs carry more lattice parameters than the standard $\Z^2$ lattice, and these parameters
do all appear in the reduced mapping. In the second case the lattice parameters are interchanged
by the shift on the quad-graph, and we find the reduction to be an alternating mapping, cf. \cite{Q}.

For certain $\s$-reductions the periodic solutions are given by
multi-valued mappings, or correspondences, see \cite{K}. The
staircase method applies equally well in such cases, see section
\ref{pkdvh}, where we perform (3,0)-reduction for the pKdV equation,
and section \ref{qdh}, where we ($n$,0)-reduce the QD-system. Here, another
question arises: {\em what is the multi-valuedness of the iterates
of the correspondence?}

In general, the number of image points of the $n$th iterate of an
$m$-valued correspondence would be $m^n$. However, it has been shown
that for completely integrable correspondences the number of images
under the iterates grows polynomially, rather than exponentially
\cite{Ves1}. In section \ref{mc} we show that for the
correspondences obtained in sections \ref{pkdvh} and \ref{qdh} the
multi-valuedness of their $n$th iterate is $n+1$ and $2n$,
respectively.

\section{The staircase method, general theory} \label{sm}
Let ${\bf u}$ be a multi-component field on the square lattice $\Z^2$ and let ${\bf
f}[{\bf u}]$ be a multi-component function of ${\bf u}$ and finitely many
shifts of ${\bf u}$.
We call a lattice equation ${\bf f}[{\bf u}]=0$ {\em integrable} if it
arises as the compatibility condition of two linear equations $\t{\psi}=L\psi$
and $\h{\psi}=M\psi$. Here, $\ \t{}\ $ denotes the horizontal shift $l\mapsto l+1$ and
$\ \h{}\ $ denotes the vertical shift $m\mapsto m+1$. Thus we have
\begin{equation} \label{lmml}
\h{L}M\equiv\t{M}L\ \text{mod}\ {\bf f},
\end{equation}
which is called the Lax-equation, or zero-curvature condition. The matrices $L$ and $M$ are
called Lax matrices.

As pointed out by Calogero and Nucci \cite{CN} in the continuum case, see also \cite{BF},
the mere existence of a Lax pair is not sufficient for integrability, the Lax pair has to be
a good Lax pair. In the discrete setting one has to be equally careful, see chapter 6 in the thesis
of Mike Hay \cite{MH}. From this point of view one could argue that the staircase method
tests whether a Lax pair is `good'. The Lax pair would be called good (and hence the lattice
equation integrable), if it can be used to produce a sufficient number of integrals for
periodic reductions.

We say that a lattice equation ${\bf f}[{\bf u}]=0$ admits a {\em
well-posed} initial value problem if from a set of generic initial
points a solution can be constructed in a unique way. An initial
value problem is called {\em nearly-well-posed} if from a set of generic initial
points solutions can be constructed, which can take only finitely many values
at each lattice point. We consider periodic initial value problems.
In the first case the solution is obtained by iterating a (finite
dimensional) mapping. In the second case the solutions are obtained by iterating
a correspondence. We note that if the initial value problem is well-posed the
periodicity of the solution is implied by the periodicity of the initial values,
whereas when in the case of nearly well-posedness the periodicity of the solutions
is imposed.

\begin{theorem} \label{ttsm}
Let $\a,\s$ be elements of $\Z\times \Z$. Suppose an integrable equation ${\bf f}[{\bf u}]=0$ allows a $\s$-periodic
initial value problem which is well-posed, or nearly-well-posed. Then, with $\cL$ being
an inversely ordered product of Lax matrices over a connected path, e.g. a staircase, from $\a$
to $\a+\s$, the trace of $\cL^i$ is invariant under any shift on the lattice, $\forall i\in\N$.
\end{theorem}

\proof
Let $\a,\b$ be two points on the lattice. Define $\cL_{\a,\b}$ to be the inversely
ordered product of Lax matrices along a connected path from $\a$ to $\b$.
We have to show that $\cL_{\a,\b}$ does not depend on the path from $\a$ to $\b$. This follows
from the fact that every square can be passed in two ways: if $L,M$ are the Lax matrices at $\a\in\Z^2$ (and
$\t{L}$ is a Lax matrix at $\t{\a}=\a+(1,0)$), then from (\ref{lmml}) it follows that
$\h{L}M=\t{M}L=\cL_{\a,\h{\t{\a}}}$ is well-defined for solutions of ${\bf f}[{\bf u}]=0$.
We have
%\begin{eqnarray*}
\[
\cL_{\t{\a},\t{\b}}
= \cL_{\b,\t{\b}}\cL_{\a,\b}\cL_{\t{\a},\a}
= \cL_{\b,\t{\b}}\cL_{\a,\b}\cL^{-1}_{\a,\t{\a}}
\]
%\end{eqnarray*}
Because the initial value problem is well-posed, or nearly-well-posed, there exists an $\s$-periodic
solution. Now let $\b=\a+i\s$ ($i\in\N$), so that the value of the solution at $\a$ and $\b$ coincide.
Then $\cL_{\a,\b}=\cL^i$. Also, $\cL_{\b,\t{\b}}=\cL_{\a,\t{\a}}$. If we denote $I=\tr(\cL^i)$,
it is clear that we have $\t{I}=I$ and, similarly $\h{I}=I$.
\qed

The mapping, or correspondence, which generates the $\s$-periodic solution
is defined by updating a set of initial values through a shift on the lattice.
Therefore, an invariant for it is given by the trace of (an integer power of) $\cL$.
If the Lax-matrices depend on a spectral parameter, say $k$, one can expand
the trace $\tr(\cL^i)$ in powers of $k$. Each coefficient then provides an integral
for the mapping, or for the correspondence. However, these integrals are not all
functionally independent.

By the Cayley-Hamilton theorem any matrix $\cL$ satisfies its own characteristic equation
$P(\lambda)=\det(\lambda I - \cL)$, i.e. we have $P(\cL)=0$. Therefore, given that $\cL$
is a $n\times n $ matrix, it suffices to consider traces of $\cL^i$, with $i\leq n$.
Even better, there are certain combinations of $\tr(\cL^i)$, which, generally, yield
a nicer basis of functionally independent integrals. These are provided by the coefficients
in $P(\lambda)=0$. For example, if $n=2$ we have
\begin{equation} \label{n=2}
P(\lambda)=\lambda^2-\lambda \tr(\cL) + (\tr(\cL)^2-\tr(\cL^2))/2.
\end{equation}
Note that the coefficient of $\lambda^0$ coincides with the determinant of $\cL$.

For general $n$, the coefficients can be obtained using Newton's identities
\begin{equation} \label{new}
ne_n=\sum_{i=1}^n (-1)^{i-1} e_{n-i} p_i,
\end{equation}
where the power sums $p_k$ are given by $p_k=x_1^k+x_2^k+\cdots+x_q^k$ %\sum_{i=1}^q x_i^k,\]
and the elementary symmetric polynomials $e_k$ are given by
\[
e_k=\sum_{1\leq i_1}\sum_{i_1<i_2}\cdots\sum_{i_{k-1}<i_k}\sum_{i_k\leq q} \prod_{j=1}^k x_{i_j},
\]
and appear as coefficients in the (Vieta) expansion
\begin{equation} \label{vie}
\prod_{i=1}^q(\lambda - x_i) =\sum_{i=0}^q (-1)^i e_i \lambda^{q-i}.
\end{equation}
If we denote the $q$ eigenvalues of the matrix $\cL$ by $x_i$, the characteristic
polynomial equals the left hand side of equation (\ref{vie}). Using Newton's identities
(\ref{new}) the right hand side can be expanded recursively in terms of $p_k=\tr(\cL^k)$.
Taking $k=1,2$ in (\ref{new}) we find familiar coefficients $e_1=p_1$, $e_2=(p_1^2-p_2)/2$,
see equation (\ref{n=2}). Taking $k=3,4$ Newton's identities yield $e_3=(p_1^3-3p_1p_2+2p_3)/6$, and
$e_4=(p_1^4-6p_1^2p_2+3p_2^2+8p_1p_3-6p_4)/24$.

For scalar equations that are defined on elementary squares, initial values are given on staircases.
So the dimension of the initial value problem is $|s_1|+|s_2|$.
A so-called standard staircase, cf. \cite{QNCL,RKQ,K} gives rise to a particularly simple mapping.
In fact, any mapping, defined by a shift on the lattice, is (equivalent to) a certain iterate of
this basic one.

\noindent
\parbox{53mm}{
\begin{center}
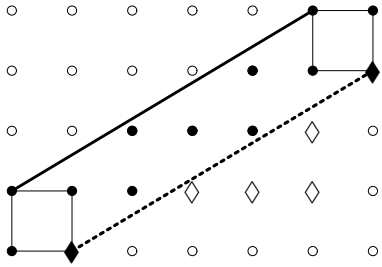

\setlength{\unitlength}{8mm}
\begin{picture}(6,4)(0,0)
\matrixput(0,2)(1,0){5}(0,1){3}{\circle{.15}}
\path(0,0)(0,1)(1,1)(1,0)(0,0)
\path(5,3)(5,4)(6,4)(6,3)(5,3)
\multiput(0,1)(2,1){4}{\circle*{.15}}
\multiput(0,0)(2,1){3}{\circle*{.15}}
\multiput(1,1)(2,1){3}{\circle*{.15}}
\put(5,4){\circle*{.15}}
\multiput(2,0)(1,0){4}{\circle{.15}}
\multiput(6,0)(0,1){3}{\circle{.15}}
\multiput(2.85,.85)(1,0){3}{$\lozenge$}
\put(4.85,1.85){$\lozenge$}
\multiput(0.85,-.15)(5,3){2}{$\blacklozenge$}
\Thicklines
\path(0,1)(5,4)
\dottedline{.14}(1,0)(6,3)
\end{picture}
\captionof{figure}{\label{s53s} (5,3)-periodic initial value problem for
equations defined on elementary squares}
\end{center}
} \quad \hfill \parbox{62mm}{To illustrate this,
we have presented the standard $(5,3)$-staircase in Figure \ref{s53s}. The standard staircase is the path
through the points between the two lines, the black dots. They are the points where initial values are given.
The standard mapping will be the shift $u_{l,m}\mapsto u_{l+2,m+1}$. Note that by this shift almost all black dots are shifted
to another black dot. The one black dot which is closest to the dotted line, is shifted to the black diamond,
whose value can be calculated using the equation on the square.}

\smallskip
\noindent
 We can also see that the mapping defined by the
shift $u_{l,m}\mapsto u_{l+1,m}$ is given as the third power of the standard one.
To evaluate the monodromy matrix $\cL$, one would take the product of matrices
along the same staircase on which the initial values are given. Since for this type of equations
the matrices $L,M$ depend on $(u,\t{u})$, $(u,\h{u})$, respectively, the matrix $\cL$ is then
automatically expressed in terms of the initial values. We note that one can just as well take the product
over any other one-period long path. For example, assuming that $\s\in\N \times \N$, one could consider the product
\[
\cL(l,m)= \prod_{j=0}^{\curvearrowleft \atop s_2-1} M_{l+s_1,m+j}
\prod_{i=0}^{\curvearrowleft \atop s_1-1} L_{l+i,m}.
\]
Then, one first has to calculate the points on the corresponding path to be able to evaluate $\cL$.
In the example given in Figure \ref{s53s}, we would need to calculate the values of the field at the
white diamonds, which amounts to iterating the mapping 8 times. %($=\max\{s_1,s_2\}$)

For equations, or systems, that are not defined on elementary squares the initial value problem
does, in general, not lie on a staircase. Depending on the type of stencil and on the particular
periodicity condition, there could either be more, or less than $|s_1|+|s_2|$ initial values.
In \cite{K} it was shown how to write down, for a given scalar equation on an arbitrary stencil,
a piece-wise linear expression (as a function of $\s$) for the dimension of an $\s$-periodic initial
value problem.

Note that the monodromy matrix is still a product of $|s_1|+|s_2|$ matrices (if
the product is taken over a staircase, which is the sensible thing to do). Also note that, in general,
the Lax matrices depend on $\u$ and a number of shifts of $\u$.
Therefore, for certain choices of $\s$, one needs to determine a number of points, by iterations
of the mapping, or correspondence, in order to evaluate the monodromy matrix in terms
of the initial values. It might be possible to avoid this by using the equation to change the
$[\u]$-dependence of the Lax-matrices. However, that would have to be adjusted to the particular
choice of $\s$.

\section{The staircase method, applications}
In this section we will apply the staircase method to a variety of different equations and different
reductions. Firstly, we consider a 5-point equation, i.e. an equation which is not defined on the square.
Second, we calculate integrals for reductions of a 2-component system of equations. We first perform
reductions which give rise to mappings and then we also perform reductions that yield correspondences.

\subsection{Mappings with a sufficient number of integrals}

\subsubsection{The Bruschi-Calogero-Droghei equation} \label{sbcd}
\noindent
\parbox{25mm}{
\setlength{\unitlength}{.8 cm}
\begin{center}
\begin{picture}(2,2)(0,0)
\matrixput(0,0)(1,0){3}(0,1){3}{\circle{.1}}
\multiput(0,0)(1,1){3}{\circle*{.05}}
\multiput(1,0)(0,2){2}{\circle*{.05}}
\path(1,0)(0,0)(2,2)(1,2)
\end{picture}
\captionof{figure}{\label{fsbcd}}
\end{center}
}
\hfill
\parbox{90mm}{In \cite{BCD} one finds the five-point equation
$E(u,\h{u},\uh{u},\t{\h{u}},\uh{\ut{u}})=0$, where
\begin{equation} \label{E}
E=(u-\uh{u})(u-\uh{\ut{u}})-(u-\h{u})(u-\t{\h{u}}),
\end{equation}
see \cite[Equation (4a)]{BCD} in which we have set $\alpha^{(\nu)}=0$. The equation is defined on the stencil given in Figure \ref{fsbcd}.}

\smallskip
\noindent
The two recursive formulas \cite[Equations (1a),(6)]{BCD}, with coefficients \cite[Equations (5),(7)]{BCD},
yield the following Lax-pair
\begin{eqnarray*}
L(u,\t{u},\uh{u},\uh{\ut{u}})&=&\left(\begin{array}{cc} k+\t{u}-u &
(u-\uh{u})(u-\uh{\ut{u}}) \\
1 & 0 \end{array} \right),\\
M(u,\h{u},\ut{u})&=&\left(\begin{array}{cc} 1 & (\h{u}-u) \\
(\h{u}-\ut{u})^{-1} & 1-(k+u-\ut{u})(\h{u}-\ut{u})^{-1} \end{array} \right).
\end{eqnarray*}
Here we have denoted the spectral parameter ($x$ in \cite{BCD}) by $k$, which we do throughout this paper.
%\noindent
%\parbox{56mm}{
%\begin{center}
%\setlength{\unitlength}{8mm}
%\begin{picture}(4,4)(0,0)
%\matrixput(0,0)(1,0){5}(0,1){5}{\circle{.15}}
%\Thicklines
%\path(1,0)(3,4)
%\path(0,2)(4,2)
%\put(3.4,2.6){$R^1$}
%\put(2.4,0.4){$R^2$}
%\put(.2,0.4){$R^1$}
%\put(1.1,2.4){$R^2$
%\end{picture}
%\captionof{figure}{\label{r5p} Distinguished regions for the Bruschi-Calogero-Droghei equation.}
%\end{center}
%} \quad \hfill \parbox{65mm}{
The method laid out in \cite{K} tells us how to pose well-defined $\s$-periodic initial
value problems for this equation.
This can be done for all $\s=(s_1,s_2)$ such that $s_2(s_2-2s_1)\neq0$.
The dimension of the periodic solutions is given by the following
piecewise-linear function
$
2\max\{|s_2-s_1|,|s_1|\}.
$
We apply the staircase method to a few reductions,
in the different regions distinguished by this function, see \cite[Figure 10]{K}.

\subsubsection*{(0,3)-reduction}
\noindent
\parbox{25mm}{
\setlength{\unitlength}{.8 cm}
\begin{center}
\begin{picture}(2,4.5)(0,-.5)
\matrixput(0,0)(1,0){2}(0,1){5}{\circle*{.1}}
\multiput(2,0)(0,1){5}{\circle{.1}}
\def\e{{.2}}
\put(-.3,-.3){$x_1$}
\put(.7,-.3){$x_2$}
\put(-.3,.7){$x_3$}
\put(.7,.7){$x_4$}
\put(-.3,1.7){$x_5$}
\put(.7,1.7){$x_6$}
\put(-.3,2.7){$x_1$}
\put(.7,2.7){$x_2$}
\put(-.3,3.7){$x_3$}
\put(.7,3.7){$x_4$}

\put(2.1,2){$\t{x}_6$}
\put(2.1,3){$\t{x}_2$}
\put(2.1,4){$\t{x}_4$}

\path(1,0)(0,0)(2,2)(1,2)
\path(1,1)(0,1)(2,3)(1,3)
\path(1,2)(0,2)(2,4)(1,4)
\end{picture}
\captionof{figure}{\label{f1}}
\end{center}
}
\hfill
\parbox{90mm}{
We assigning initial values as in Figure \ref{f1}. They are indicated by the black dots. We update them
using the right-shift, the values of $\t{x_2},\t{x_4},\t{x_6}$ can be determined using
the equations indicated by the zig-zags. We get a six-dimensional mapping
\[
\left( \begin{array}{cc}
x_1 & x_2 \\
x_3 & x_4 \\
x_5 & x_6
\end{array}
\right)
\mapsto
\left(
\begin{array}{cc}
x_2 & \frac{x_2x_6+x_3x_4-x_3x_6-x_4x_6}{x_2-x_6} \\
x_4 & \frac{x_2x_5+x_2x_6-x_2x_4-x_5x_6}{x_2-x_4}\\
x_6 & \frac{x_1x_4+x_2x_4-x_1x_2-x_4x_6}{x_4-x_6}
\end{array}
\right) .
\]
The monodromy matrix, which we take from $x_2$ upwards to $x_2$, is $
\cL= M(x_6,x_2,x_5)M(x_4,x_6,x_3)M(x_2,x_4,x_1).
$}

\smallskip
\noindent
Three functionally independent integrals for this mapping can be obtained from the coefficients in its characteristic polynomial (\ref{n=2}).
They are $(x_1-x_4)(x_2-x_5)(x_3-x_6)$, $(x_1-x_6)(x_2-x_3)(x_4-x_5)$, and
$(x_1 - x_5) (x_2 - x_4) + (x_3-x_5) (x_4 - x_6 )$.

\subsubsection*{(1,3)-reduction}
\noindent
\parbox{25mm}{
\setlength{\unitlength}{.8 cm}
\begin{center}
\begin{picture}(2,3)(0,-.5)
\matrixput(0,0)(1,0){3}(0,1){4}{\circle{.1}}
\multiput(1,0)(0,1){4}{\circle*{.1}}
\multiput(0,0)(2,3){2}{\circle*{.1}}
\put(-.3,-.3){$x_1$}
\put(.7,-.3){$x_4$}
\put(.7,.7){$x_3$}
\put(.7,1.7){$x_2$}
\put(.7,2.7){$x_1$}
\put(1.7,2.7){$x_4$}
\path(1,0)(0,0)(2,2)(1,2)
\put(2.1,2){$\uh{x_4}$}
\end{picture}
\captionof{figure}{\label{f2}}
\end{center}
}
\hfill
\parbox{90mm}{
Assigning initial values as in Figure \ref{f2}, and updating them
using the down-shift, we get a four-dimensional mapping
\begin{equation} \label{4dm}
(x_1,x_2,x_3,x_4) \mapsto (x_2 , x_3 , x_4 , \frac{x_1x_4+x_2x_3-x_1x_3-x_3x_4}{x_2-x_3} ).
\end{equation}
In this case we have to first calculate a few values of the field at points close to the staircase
in order to evaluate the monodromy matrix. We calculate $\h{x_1}$ by solving $E(x_2,x_1,x_3,x_4,\h{x_1})=0$, and
we find $\h{\h{x_1}}$ from $E(x_1,\h{x_1},x_2,x_3,\h{\h{x_1}})=0$.}

\smallskip
\noindent
Two functionally independent integrals are obtained from the coefficients
in (\ref{n=2}) with
$
\cL=M(x_2,x_1,\h{\h{x}}_1)M(x_3,x_2,\h{x}_1)M(x_4,x_3,x_1)L(x_1,x_4,x_2,\h{\h{x}}_1).
$
They are $J_1=(x_1-x_2)(x_3-x_4)$, and $J_2=(x_1-x_3)(x_2-x_4)/(x_2-x_3)$.

\subsubsection*{(2,3)-reduction}
\noindent
\parbox{25mm}{
\setlength{\unitlength}{.8 cm}
\begin{center}
\begin{picture}(2,3)(0,0)
\matrixput(0,0)(1,0){3}(0,1){4}{\circle{.1}}
\multiput(0,0)(1,3){2}{\circle*{.1}}
\multiput(0,1)(1,1){3}{\circle*{.1}}
\put(-.3,-.3){$x_4$}
\put(-.3,.7){$x_2$}
\put(.7,1.7){$x_3$}
\put(.7,2.7){$x_1$}
\put(1.7,2.7){$x_4$}
\put(1.1,1){$\h{\t{x_4}}$}
\path(1,1)(0,1)(2,3)(1,3)
\end{picture}
\captionof{figure}{\label{f3}}
\end{center}
}
\hfill
\parbox{90mm}{We assign initial values as in Figure \ref{f3} and update them
using the diagonal shift $u\mapsto\h{\t{u}}$. The values $\t{x}_2$, which equals $\h{\t{x_4}}$, and $\t{x}_3$ are
determined by $E(x_3,x_1,\t{x}_2,x_4,x_2)=0$, and $E(x_4,x_2,\t{x}_3,\t{x}_2,x_3)=0$,
successively. We find the same four-dimensional mapping as in the previous case.
The monodromy matrix $M(x_4,x_2,x_1)${}$M(\t{x}_3,x_4,x_3)${}$L(x_3,\t{x}_3,\t{x}_2,x_2)
${}$M(\t{x}_2,x_3,x_2)\cdot$ $L(x_2,\t{x}_2,x_4,x_1)$ yields the same integrals $J_1,J_2$.
}

\subsubsection*{(2,-1)-reduction}
\noindent
\parbox{25mm}{
\setlength{\unitlength}{.8 cm}
\begin{center}
\begin{picture}(2,3)(0,0)
\matrixput(0,0)(1,0){3}(0,1){4}{\circle{.1}}
\matrixput(0,1)(1,0){2}(0,1){3}{\circle*{.1}}
\multiput(2,0)(0,1){3}{\circle*{.1}}
\multiput(-.3,.7)(2,-1){2}{$x_1$}
\multiput(-.3,1.7)(2,-1){2}{$x_3$}
\multiput(-.3,2.7)(2,-1){2}{$x_5$}
\put(.7,.7){$x_2$}
\put(.7,1.7){$x_4$}
\put(.7,2.7){$x_6$}
\path(1,1)(0,1)(2,3)(1,3)
\put(2.1,3){$\t{x_6}$}
\end{picture}
\captionof{figure}{\label{f4}}
\end{center}
}
\hfill
\parbox{90mm}{Assigning initial values as in Figure \ref{f4}, and updating them
using the right-shift, we get a six-dimensional mapping
\begin{align*}
x_i &\mapsto x_{i+1},\quad i\in\{1,2,\ldots 5\}, \\
x_6 &\mapsto \frac{x_1x_4+x_2x_4-x_1x_2-x_4x_6}{x_4-x_6}.
\end{align*}
The trace of the monodromy matrix
$M^{-1}(x_4,x_6,x_3)L(x_5,x_6,x_3,x_2)L(x_4,x_5,x_2,x_1)$ yields two functionally independent
integrals}

\smallskip
\noindent
These are $(x_1-x_6)(x_2-x_4)(x_3-x_5)$ and $(x_2-x_6)(x_3-x_5)+(x_1-x_4)(x_2-x_4)+(x_3-x_4)(x_4-x_5$.
A third functionally independent integral, $(x_1-x_4)(x_2-x_4)(x_2-x_5)(x_3-x_5)(x_3-x_6)$, is obtained by taking
the determinant of the monodromy matrix. In the previous cases, the determinant
does not provide a functionally independent integral.

\subsubsection{The QD-algorithm} \label{sqd}
\noindent
\parbox{25mm}{
\setlength{\unitlength}{.8cm}
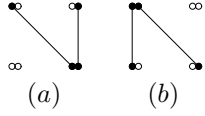
\begin{center}
\begin{picture}(4,1.5)(-.3,-1)
\matrixput(-.05,0)(1,0){4}(0,1){2}{\circle{.10}}
\matrixput(.05,0)(1,0){4}(0,1){2}{\circle{.10}}
\multiput(-.05,1)(2,0){2}{\circle*{.1}}
\multiput(.95,0)(1,0){2}{\circle*{.1}}
\multiput(1.05,0)(2,0){2}{\circle*{.1}}
\multiput(1.05,1)(1,0){2}{\circle*{.1}}
\path(-.05,1)(.95,0)(1.05,0)(1.05,1)
\path(1.95,0)(1.95,1)(2.05,1)(3.05,0)
\put(.2,-.6){($a$)}
\put(2.2,-.6){($b$)}
\end{picture}
\captionof{figure}{\label{fqd} \\ QD-type system}
\end{center}
} \quad \hfill \parbox{90mm}{
The quotient-difference (QD) algorithm,
\begin{align}
\h{e} + \h{\t{q}} &= \t{q} + \t{e}, \tag{11a} \\
\h{e} \h{q} &= e \t{q}, \tag{11b}
\addtocounter{equation}{1}
\end{align}
is used to construct continued fractions whose convergents form
ordered sequences in a normal Pad\'{e} table \cite{Cabe}, and to
find the zeros of a polynomial \cite{HW}.}

\smallskip
\noindent
It is also called the time-discrete Toda molecule \cite{NTS}. It is an integrable two-component equation
%of the form
%\begin{equation} \label{tqd}
%f(\h{e},\t{e},\t{q},\h{\t{q}})=0,\quad
%g(e,\h{e},\t{q},\h{q})=0,
%\end{equation}
defined on the stencils depicted in Figure \ref{fqd}, where we associate two values, $e$ on the left and and $q$ on the right, to every point on the lattice.

A Lax-pair for the QD-algorithm can be obtained from relations between so called higher adjacent orthogonal polynomials \cite{Brez}, cf.
\cite[equations (3,4)]{PGR}
\[
k\h{P}=\t{P}+qP, \quad \t{P}=\h{\t{P}}+e\h{P}.
\]
With $\Psi^t=(P,\h{P})$ we have $\t{\Psi}=L \Psi$ and
$\h{\Psi}=k^{-1} M \Psi$, where
\[
L(e,q) = \left( \begin{array}{cc} -q & k \\ -q & k-e \end{array} \right),
\qquad M(e,q,\h{q}) = \left( \begin{array}{cc} 0 & k \\ -q & k-e+\h{q}
\end{array} \right).
\]
This (small) Lax-pair differs from the (big) Lax-pair obtained in
\cite[equation (9)]{PGR}. A big Lax-pair incorporates a particular
choice of initial values and we would like to consider general
periodic initial value problems.\\

\noindent
\parbox{40mm}{
\setlength{\unitlength}{8mm}
\begin{center}
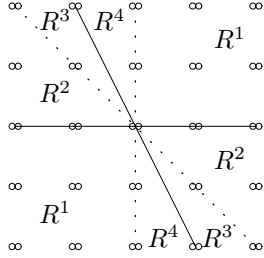

\begin{picture}(4,4)(0,0)
\matrixput(-0.05,0)(1,0){5}(0,1){5}{\circle{.1}}
\matrixput(0.05,0)(1,0){5}(0,1){5}{\circle{.1}}
\path(1,4)(3,0)
\dottedline{.25}(0,4)(4,0)
\path(0,2)(4,2)
\dottedline{.25}(2,0)(2,4)
\put(3.3,3.3){$R^1$}
\put(3.3,1.3){$R^2$}
\put(3.1,0){$R^3$}
\put(2.2,0){$R^4$}
\put(.4,.4){$R^1$}
\put(.4,2.4){$R^2$}
\put(.4,3.6){$R^3$}
\put(1.3,3.6){$R^4$}
\end{picture}
\captionof{figure}{\label{rqd} Distinct regions for the QD-stencil}
\end{center}
} \quad \hfill \parbox{75mm}{
In \cite{K}, it has been shown that for all $s=(s_1,s_2)\in\Z \times \Z$ such that
$s_2(s_2-2s_1)\neq0$, there exists a well-posed $\s$-periodic
initial value problem, with dimension
\[
2\max(|s_1+s_2|,|s_1|).
\]
This function tells us there are two different regions, where the dimension is
given by a different linear function of the periods (up to a sign).
However, to pose the initial value problems
one has to distinguish four different regions, as depicted in Figure \ref{rqd}.}

\smallskip
\noindent
We note that for all $\s$ initial values can be given on (part of) a standard staircase. In the Figures that follow this will be indicated by a dotted line. We will present four examples of families of periodic reductions, where the dimension depends on an arbitrary variable $n\in\N$, one family in each of the different regions.

\subsubsection*{($0,n$)-reduction}
\noindent
\parbox{40mm}{
\setlength{\unitlength}{12mm}
\begin{center}
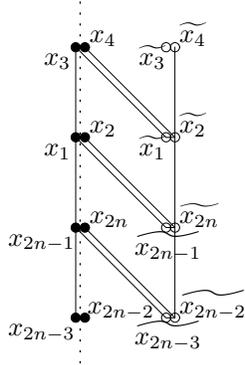

\begin{picture}(2,4)(-.5,0)
\multiput(.95,.5)(0,1){4}{\circle{.10}}
\multiput(-.05,.5)(0,1){4}{\circle*{.10}}
\multiput(1.05,.5)(0,1){4}{\circle{.10}}
\multiput(0.05,.5)(0,1){4}{\circle*{.10}}
\dottedline{.12}(0,0)(0,4)
\put(-.8,.3){$x_{2n-3}$}
\put(.08,.55){$x_{2n-2}$}
\put(-.8,1.3){$x_{2n-1}$}
\put(.1,1.55){$x_{2n}$}
\put(-.4,2.3){$x_1$}
\put(.1,2.55){$x_2$}
\put(-.4,3.3){$x_3$}
\put(.1,3.55){$x_4$}
\put(.6,.2){$\t{x_{2n-3}}$}
\put(1.1,.55){$\t{x_{2n-2}}$}
\put(.6,1.2){$\t{x_{2n-1}}$}
\put(1.1,1.55){$\t{x_{2n}}$}
\put(.65,2.3){$\t{x_1}$}
\put(1.1,2.55){$\t{x_2}$}
\put(.65,3.3){$\t{x_3}$}
\put(1.1,3.55){$\t{x_4}$}
\path(-.05,1.5)(.95,0.5)(1.05,0.5)(1.05,1.5)
\path(-.05,0.5)(-.05,1.5)(.05,1.5)(1.05,0.5)
\path(-.05,2.5)(.95,1.5)(1.05,1.5)(1.05,2.5)
\path(-.05,1.5)(-.05,2.5)(.05,2.5)(1.05,1.5)
\path(-.05,3.5)(.95,2.5)(1.05,2.5)(1.05,3.5)
\path(-.05,2.5)(-.05,3.5)(.05,3.5)(1.05,2.5)
\end{picture}
\captionof{figure}{\label{qd0n} Vertical periodicity.}
\end{center}
}\quad \hfill \parbox{75mm}{
We take initial values $(e,q)_{0,i}=(x_{2i-1},x_{2i})$, with
$x_k=x_m$ if $k\equiv m \mod 2n$, see Figure \ref{qd0n}.

Updating these values to the right gives a $2n$-dimensional volume-preserving mapping, with $i=1,2,\ldots,2n$,
\begin{align*}
x_{2i-1} &\mapsto x_{2i+1}+\frac{x_{2i+3}x_{2i+4}}{x_{2i+1}}-\frac{x_{2i+1}x_{2i+2}}{x_{2i-1}}, \\
x_{2i} &\mapsto \frac{x_{2i+1}x_{2i+2}}{x_{2i-1}}.
\end{align*}
Here we first used equation ($11b$) to find the images $\t{x_{2i}}$, and then equation ($11a$) to find $\t{x_{2i-1}}$.
The monodromy matrix is
\begin{align*}
\cL=&M(x_{2n-1},x_{2n},x_2)M(x_{2n-3},x_{2n-2},x_{2n})\cdots \\
&\cdots M(x_3,x_4,x_6)M(x_1,x_2,x_4).
\end{align*}}

\smallskip
\noindent
 We have verified up to $n=9$ that the coefficients in the $k$-expansions of its trace and determinant
yield $n+1$ functionally independent integrals. The lowest non-trivial mapping, taking $n=2$, is
\[
(x_1, x_2, x_3, x_4) \mapsto \left( x_3+\frac{x_1x_2}{x_3}-\frac{x_3x_4}{x_1} , \frac{x_3x_4}{x_1} ,
 x_1+\frac{x_3x_4}{x_1}-\frac{x_1x_2}{x_3} , \frac{x_1x_2}{x_3}\right),
\]
which admits the three functionally independent integrals
$x_1+x_3,x_2x_4,x_1x_2+x_3x_4-x_1x_3$.

\subsubsection*{($3n,-2n$)-reduction}
We pose initial values as in Figure \ref{qd3-2} and update then using a horse-jump $(l,m) \mapsto (l+2,m-1)$. Note that one should first update $x_{6i+2}$ before $x_{6i+4}$.

\smallskip
\noindent
\parbox{60mm}{
\setlength{\unitlength}{8mm}
\begin{center}
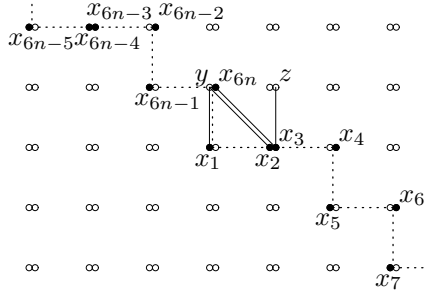

\begin{picture}(6,4.5)(0,0)
\matrixput(-.05,0)(1,0){7}(0,1){5}{\circle{.1}}
\matrixput(0.05,0)(1,0){7}(0,1){5}{\circle{.1}}
\multiput(-0.05,4)(3,-2){3}{\circle*{.1}}
\multiput(.95,4)(3,-2){2}{\circle*{.1}}
\multiput(1.05,4)(3,-2){2}{\circle*{.1}}
\multiput(2.05,4)(3,-2){2}{\circle*{.1}}
\multiput(1.95,3)(3,-2){2}{\circle*{.1}}
\multiput(3.05,3)(3,-2){2}{\circle*{.1}}
\dottedline{.12}(0,4.5)(0,4)(2,4)(2,3)(3,3)(3,2)(5,2)(5,1)(6,1)(6,0)(6.5,0)
\put(.9,4.2){$x_{6n-3}$}
\put(2.1,4.2){$x_{6n-2}$}
\put(3.1,3.1){$x_{6n}$}
\put(4.1,2.1){$x_{3}$}
\put(5.1,2.1){$x_{4}$}
\put(6.1,1.1){$x_{6}$}
\put(-.4,3.7){$x_{6n-5}$}
\put(.7,3.7){$x_{6n-4}$}
\put(1.7,2.7){$x_{6n-1}$}
\put(2.7,1.7){$x_{1}$}
\put(3.7,1.7){$x_{2}$}
\put(4.7,.7){$x_{5}$}
\put(5.7,-.3){$x_{7}$}
\put(2.7,3.1){$y$}
\put(4.1,3.1){$z$}
\path(2.95,3)(3.95,2)(4.05,2)(4.05,3)
\path(2.95,2)(2.95,3)(3.05,3)(4.05,2)
\end{picture}
\captionof{figure}{\label{qd3-2} Initial values on part of a standard ($3n,-2n$)-staircase, and the image points $y,z$ of $x_{6n-4},x_{6n-2}$.}
\end{center}
} \hfill \parbox{55mm}{
Thus we get a $6n$-dimensional mapping, with $i=0,1,\ldots,n-1$,
\begin{align*}
x_{6i+1} & \mapsto x_{6i+5}, \\
x_{6i+2} & \mapsto \frac{x_{6i+7}x_{6i+9}}{x_{6i+6}}, \\
x_{6i+3} & \mapsto x_{6i+6}, \\
x_{6i+4} & \mapsto x_{6i+8}+x_{6i+9}(1-\frac{x_{6i+7}}{x_{6i+6}}), \\
x_{6i+5} & \mapsto x_{6i+8}, \\
x_{6i+6} & \mapsto x_{6i+10},
\end{align*}
where the subscript on $x$ is taken modulo $6n$.}

\smallskip
\noindent
This mapping is measure-preserving with density $(\prod_{i=1}^n x_{6i-3})^{-1}$.
We obtain integrals by expanding the trace and determinant of the monodromy matrix
\begin{align*}
\prod_{i=0}^{\curvearrowleft \atop {n-1}} M^{-1}(x_{6i+7},x_{6i+5}+x_{6i+6}-x_{6i+7},x_{6i+6}) L(x_{6i+5},x_{6i+2}+x_{6i+4}-x_{6i+5}) \\
\cdot M^{-1}(x_{6i+5},x_{6i+2}+x_{6i+4}-x_{6i+5},x_{6i+4}) L(x_{6i+2},x_{6i+3}) \\
\cdot L(x_{6i+1},x_{6i-1}+x_{6i}-x_{6i+1})
\end{align*}
in powers of the spectral parameter $k$. We verified that, up to $n=3$, $3n+1$ of them
are functionally independent. For $n=1$ the mapping reads
\[
(x_1, x_2, x_3, x_4, x_5, x_6) \mapsto
\left( x_5 , \frac{x_1x_3}{x_6} , x_6 , x_2+x_3(1-\frac{x_1}{x_6}) , x_2 , x_4\right),
\]
which admits the following four functionally independent integrals
\[
x_6+x_4+x_2+x_3, x_3(x_4-x_5)(x_1-x_6) ,x_3(x_6+x_4-x_1)+x_6(x_4+x_2-x_5),x_1x_2x_3x_5.
\]

\subsubsection*{($2n,-3n$)-reduction}
\noindent
\parbox{35mm}{
\setlength{\unitlength}{8mm}
\begin{center}
\begin{picture}(4,6)(0,0)
\matrixput(.05,0)(1,0){5}(0,1){7}{\circle{.1}}
\matrixput(-.05,0)(1,0){5}(0,1){7}{\circle{.1}}
\multiput(-.05,6)(2,-3){3}{\circle*{.1}}
\multiput(1.05,6)(2,-3){2}{\circle*{.1}}
\multiput(.95,5)(2,-3){2}{\circle*{.1}}
\multiput(2.05,4)(2,-3){2}{\circle*{.1}}
\dottedline{.12}(0,6.5)(0,6)(1,6)(1,5)(2,5)(2,3)(3,3)(3,2)(4,2)(4,0)(4.5,0)
\put(-.3,5.7){$x_{4n-3}$}
\put(1,6.2){$x_{4n-2}$}
\put(.7,4.7){$x_{4n-1}$}
\put(2.1,4.1){$x_{4n}$}
\put(1.7,2.7){$x_{1}$}
\put(3.1,3.1){$x_{2}$}
\put(2.7,1.7){$x_{3}$}
\put(4.1,1.1){$x_{4}$}
\put(3.7,-.3){$x_{5}$}
\path(0.95,5)(1.95,4)(2.05,4)(2.05,5)
\path(1.95,3)(1.95,4)(2.05,4)(3.05,3)
\end{picture}
\captionof{figure}{\label{2-3} Part of the ($2n,-3n$)-staircase}
\end{center}
} \quad \hfill \parbox{80mm}{
We choose initial values as in Figure \ref{2-3},
\begin{align*}
e_{2i,-3i}=x_{4i+1}&, \quad q_{2i+1,-3i}=x_{4i+2} \\
e_{2i+1,-3i-1}=x_{4i+3}&, \quad q_{2i+2,-3i-2}=x_{4i+4},
\end{align*}
with $x_k=x_m$ if $k\equiv m\ \mathrm{mod}\ 4n$. They are updated by shifting $(l,m) \mapsto (l+1,m-1)$. This yields the $4n$-dimensional mapping
\begin{align*}
x_{4i+1} & \mapsto x_{4i+3}, \\
x_{4i+2} & \mapsto \frac{x_{4i+5}x_{4i+6}}{x_{4i+4}}+x_{4i+4}-x_{4i+3}, \\
x_{4i+3} & \mapsto \frac{x_{4i+5}x_{4i+6}}{x_{4i+4}}, \\
x_{4i+4} & \mapsto x_{4i+6},
\end{align*}
which is measure-preserving with density $\prod_{i=1}^n x_{4i}$.
}

\smallskip
\noindent
The monodromy matrix is
\begin{align*}
\prod_{i=0}^{\curvearrowleft \atop {n-1}} M^{-1}(x_{4i+5},p_i,x_{4i+4})
&M^{-1}(e_i,x_{4i+4},q_i)L(x_{4i+3},r_i) \\
&\cdot M^{-1}(x_{4i+3},r_i,x_{4i+2})L(x_{4i+1},p_{i-1}),
\end{align*}
where $e_i=x_{4i+5}x_{4i+6}/x_{4i+4}$, $ q_i=e_i+x_{4i+4}-x_{4i+3}$, $r_i=x_{4i+1}+x_{4i+2}-x_{4i+3}$,
$p_i=z_i+x_{4i+4}-x_{4i+5}$, and $z_i=x_{4i+3}r_i/x_{4i+4}$. We verified up to $n=3$ that its trace and
determinant yield $2n+1$ functionally independent integrals. For $n=1$ the measure preserving 4-dimensional
mapping reads
\begin{equation} \label{sam}
(x_1,x_2,x_3,x_4) \mapsto
\left(x_3 , \frac{x_1x_2}{x_4}+x_4-x_3 , \frac{x_1x_2}{x_4} , x_2\right),
\end{equation}
which admits the following three functionally independent integrals,
\[
x_2+x_4-x_3,\ \frac{x_1x_3}{x_4},\ \frac{(x_1-x_4)(x_2-x_3)(x_3-x_4)}{x_4}.
\]

\subsubsection*{($1,-1-n$)-reduction}
\noindent
\parbox{25mm}{
\setlength{\unitlength}{8mm}
\begin{center}
\begin{picture}(2,6.5)(-1,0)
\matrixput(.05,0.5)(1,0){2}(0,1){6}{\circle{.1}}
\matrixput(-.05,0.5)(1,0){2}(0,1){6}{\circle{.1}}
\multiput(.95,.5)(0,1){2}{\circle*{.10}}
\multiput(-.05,3.5)(0,1){3}{\circle*{.10}}
\multiput(1.05,.5)(0,1){3}{\circle*{.10}}
\multiput(0.05,4.5)(0,1){2}{\circle*{.10}}
\dottedline{.12}(0,6)(0,3.5)(1,3.5)(1,.5)(1,0)
\put(-.6,3.6){$x_1$}
\put(-.6,4.6){$x_2$}
\put(-.6,5.6){$x_3$}
\put(.1,4.65){$x_{n+1}$}
\put(.1,5.65){$x_{n+2}$}
\put(.0,.6){$x_{n-1}$}
\put(.4,1.6){$x_{n}$}
\put(1.1,.6){$x_{2n-2}$}
\put(1.1,1.6){$x_{2n-1}$}
\put(1.1,2.6){$x_{2n}$}
\path(-.05,3.5)(.95,2.5)(1.05,2.5)(1.05,3.5)
\path(-.05,3.5)(-.05,4.5)(.05,4.5)(1.05,3.5)
\end{picture}
\captionof{figure}{\label{qd1-n-1}}
\end{center}
} \quad \hfill \parbox{90mm}{
We choose initial values as in Figure \ref{qd1-n-1}, with
$j=1,2,\ldots,n$, and $i\in\Z$,
\[
e_{i,-i(n+1)+j-1}=x_j,\  q_{i,-i(n+1)+j}=x_{n+j}.
\]
Updating by the up-shift yields a $2n$-dimensional mapping
\begin{align*}
x_{i} & \mapsto x_{i+1} ,\quad i \in \{1,2,\ldots,2n-1\}, i\neq n \\
x_n & \mapsto x_1+\frac{x_2x_{n+1}}{x_1}-x_{2n}, \\
x_{2n} & \mapsto \frac{x_2x_{n+1}}{x_1},
\end{align*}
which is measure-preserving with density $x_1$.
}

\smallskip
\noindent
The monodromy matrix is
\begin{align*}
M^{-1}&(x_1,(x_n+x_{2n-1}-x_{2n})\frac{x_{2n}}{x_1},x_{n+1})
(\prod_{i=2}^n M^{-1}(x_i,x_{n+i-1},x_{n+i})) \\
&\cdot M^{-1}(x_1+\frac{x_2x_{n+1}}{x_1}-x_{2n},x_{2n},\frac{x_2x_{n+1}}{x_1})
L(x_1,(x_n+x_{2n-1}-x_{2n})\frac{x_{2n}}{x_1}),
\end{align*}
whose trace and determinant yield $n+1$ functionally independent integrals,
which we verified up to $n=8$. For $n=2$ we find mapping (\ref{sam}) again,
under the change of variables
\[
(x_1,x_2,x_3,x_4) \mapsto (x_4,x_2,x_1,x_3).
\]

\subsection{Correspondences with a sufficient number of integrals}
There are certain lines in the ($s_1,s_2$)-plane where a periodic
reduction yields a correspondence instead of a mapping, see \cite{K}.
Here we impose the solution to be periodic, which is not
implied by the periodicity of the initial conditions.
For the QD-system, we find correspondences
on lines given by $s_2(s_2+2s_1)=0$, cf. \cite{K}.

\subsubsection{The QD-algorithm, ($n,0$)-reduction} \label{qdh}
We consider horizontal staircases for the QD-system.
Solving the non-local (or, implicit) scheme we find rational expressions for two-valued correspondences.

\noindent
\parbox{50mm}{
\setlength{\unitlength}{12mm}
\begin{center}
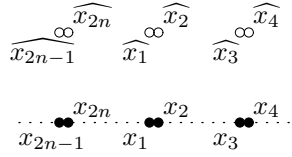

\begin{picture}(2,2)(0,-.5)
\multiput(-.05,1)(1,0){3}{\circle{.1}}
\multiput(-.05,0)(1,0){3}{\circle*{.1}}
\multiput(0.05,1)(1,0){3}{\circle{.1}}
\multiput(0.05,0)(1,0){3}{\circle*{.1}}
\dottedline{.12}(-.5,0)(2.5,0)
\put(.1,0.1){$x_{2n}$}
\put(1.1,0.1){$x_{2}$}
\put(2.1,0.1){$x_{4}$}
\put(-.5,-0.25){$x_{2n-1}$}
\put(0.65,-0.25){$x_{1}$}
\put(1.65,-0.25){$x_{3}$}
\put(.1,1.1){$\h{x_{2n}}$}
\put(1.1,1.1){$\h{x_{2}}$}
\put(2.1,1.1){$\h{x_{4}}$}
\put(-.6,0.7){$\h{x_{2n-1}}$}
\put(0.65,0.7){$\h{x_{1}}$}
\put(1.65,0.7){$\h{x_{3}}$}
\end{picture}
\captionof{figure}{Initial values and their images under the up-shift. \label{qdhf}}
\end{center}
} \hfill \parbox{65mm}{
As initial conditions we take
\[
e_{m,0}=x_{2m-1}, \quad q_{m,0}=x_{2m},
\]
where the index on $x$ is taken modulo $2n$, see Figure \ref{qdhf}. They are updated by the up-shift.
We assume the image is periodic with the same period as the initial values, i.e. we also take the index on $\h{x}$ modulo $2n$.}

\smallskip
\noindent
We have the
following $2n$ equations for the $2n$ unknowns $\h{x_i}$ (the reader might like to draw a few of them into Figure 13)
\begin{equation} \label{deqs}
\h{x_{2i-1}}\h{x_{2i}}=x_{2i-1}x_{2i+2}, \quad
\h{x_{2i-1}} + \h{x_{2i+2}} = x_{2i+1} + x_{2i+2}.
\end{equation}
We first solve for the odd variables, thereby obtaining a set of $n$
equations for the even variables
\[
\h{x_{2i-1}}=\frac{x_{2i-1}x_{2i+2}}{\h{x_{2i}}} = x_{2i+1} + x_{2i+2} - \h{x_{2i+2}}.
\]
We write $\h{x_{2i+2}} = m_{2i+2}(\h{x_{2i}})$, where
\[
m_k(z)=x_{k-1} + x_{k} - \frac{x_{k-3}x_{k}}{z}.
\]
Now $\h{x_{2k}}$ must be one of the two fixed points of the M\"{o}bius transformation
\begin{align*}
M_{k}&=m_{2k}m_{2(k-1)}m_{2(k-2)}\cdots m_{2(k-n+1)} \\
&=m_{2k}\cdots m_{2(k+3)}m_{2(k+2)} m_{2(k+1)}.
\end{align*}
The first fixed point of $M_{k}$ is given by $x_{2k-1}$, as $m_i(x_{i-3})=x_{i-1}$.
This gives us one way of updating our initial values, i.e. the linear map $\Pm_{2n}:$
\begin{equation} \label{Pm}
\begin{split}
x_{2i-1} &\mapsto x_{2i+2}, \\
x_{2i} &\mapsto x_{2i-1},
\end{split}
\end{equation}
taking $i=1,2,\ldots,n$, assuming the index on $x$ to be periodic modulo $2n$.

The other fixed point of $M_{2k}$ is $z_{k}:=x_{2(k+1)}Q^n_{2k}/Q^n_{2k-2}$,
where
\begin{equation} \label{Q}
Q^n_k(x)=\sum_{i=1}^{n-1} \left( \prod_{j=1}^{i-1} x_{2j+2+k} \right) \left( \prod_{j=i}^{n-1} x_{2j+1+k} \right) .
\end{equation}
This follows from
\begin{lemma}
\[
m_{2k}(z_{k-1})=z_{k}
\]
\end{lemma}
\proof
In terms of $Q$ the statement is
\begin{equation} \label{lq}
x_{2k-3}Q^n_{2k-4}+x_{2k+2}Q^n_{2k}=(x_{2k-1}+x_{2k})Q^n_{2k-2}
\end{equation}
From the definition (\ref{Q}) it follows that
\begin{equation} \label{rq}
x_{2n-1}Q^{n-1}_0 + \prod_{j=2}^n x_{2j} \ = \ Q^n_0 \ = \
x_{4}Q^{n-1}_2 + \prod_{j=2}^n x_{2j-1}.
\end{equation}
Therefore we have
\begin{align*}
&x_{2n-1} Q^n_{-2} + x_4 Q_2 \\
&=x_{2n-1} (x_2 Q^{n-1}_0 + \prod_{j=1}^{n-1} x_{2j-1}) + x_4 ( x_1 Q_2^{n-1} + \prod_{j=3}^{n+1} x_{2j}) \\
&=x_2 ( x_{2n-1} Q^{n-1}_0 + \prod_{j=2}^{n} x_{2j} ) + x_1 ( x_4 Q_2^{n-1} + \prod_{j=2}^n x_{2j-1} )\\
&=(x_1+x_2) Q^n_0,
\end{align*}
which is equation (\ref{lq}) with $k=1$. This implies that equation (\ref{lq}) holds for all $k$, as we may
shift $x_i \mapsto x_{i+2(k-1)}$.
\qed

\bigskip
The mapping that corresponds to the fixed point $z_{k}$ will be denoted $\Rm_{2n}:$
\begin{equation} \label{Rm}
\begin{split}
x_{2i-1} &\mapsto x_{2i-1} \frac{Q^n_{2i-2}}{Q^n_{2i}}, \\
x_{2i} &\mapsto x_{2i+2} \frac{Q^n_{2i}}{Q^n_{2i-2}}.
\end{split}
\end{equation}
Since a M\"{o}bius transformation has at most two fixed points we obtained a two-valued
correspondence $(\Pm,\Rm)$. Integrals for this correspondence are given by the
coefficients of the $k$-expansions of the trace and determinant of the
monodromy matrix
\[
L(x_{2n},x_{2n-1})\cdots L(x_4,x_3)L(x_2,x_1).
\]
For all $n<10$
we found $n+1$ functionally independent integrals. The lowest
non-trivial case is $n=2$. Explicitly, both mappings $\Pm_4:$
\[
(x_1,x_2,x_3,x_4)\mapsto (x_4,x_1,x_2,x_3)
\]
and $\Rm_4:$
\[
(x_1,x_2,x_3,x_4) \mapsto \left( x_1\frac{x_3+x_4}{x_1+x_2} , x_4\frac{x_1+x_2}{x_3+x_4},
 x_3\frac{x_1+x_2}{x_3+x_4} , x_2\frac{x_3+x_4}{x_1+x_2} \right)
\]
admit the three invariants $x_1+x_2+x_3+x_4$, $x_1x_3+x_2x_4$, and $x_1x_2x_3x_4$.

\section{Reduction of order} \label{dr}
In all cases we have encountered so far, the staircase method
provided a sufficient number of integrals, that is, at least half the
dimension of the mapping. Still, the dimension of those mappings may
be reduced. This can be done using a symmetry of the lattice
equation. In certain cases the number of invariants remains the
same, whereas in other cases it drops.

For reductions of other equations, which we encounter in this
section, the number of integrals provided by the staircase method
is not sufficient. However, in
these cases, after reduction of order the number of integrals
will suffice. We observe that the number of dimensions to be reduced
varies with the period $\s$. This can be understood exploiting
symmetries of the P$\Delta$E that give rise, for certain periods, to
$k$-symmetries of the mappings.

\subsection{Mappings with sufficiently many integrals, revisited}

\subsubsection{The Bruschi-Calogero-Droghei equation}
For the mappings obtained in section \ref{sbcd} the number of
functionally independent integrals is exactly half the dimension of
the mapping, sufficiently many for complete integrability. We note
that the equation $E=0$, cf. (\ref{E}), admits two Lie-point symmetries $u\mapsto
u+\epsilon$ and $u\mapsto \lambda u$. This yields two symmetries for
the mappings, which can be used to reduce the dimension of the
mappings by 2. The integrals we have given only admit the first
(translation) symmetry. However, certain homogeneous combinations of
them also admit the second (scaling) symmetry. Therefore, applying
$2$-reduction to the examples in section \ref{sbcd}  produces
2-, respectively 4-dimensional mappings with 1, respectively 2 integrals.
For instance, using reduced variables
\[
z_1=\frac{x_1-x_2}{x_2-x_3},\quad z_2=\frac{x_2-x_3}{x_3-x_4},
\]
the 4-dimensional mapping (\ref{4dm}) reduces to
\begin{equation} \label{I4}
(z_1,z_2) \mapsto \left(z_2,\frac{1}{z_1}\right),
\end{equation}
which has one integral
\[
\frac{J_2^2}{J_1}=\frac{(z_1+1)^2(z_2+1)^2}{z_1z_2}.
\]
Note that the 4th iterate of (\ref{I4}) equals the identity and hence the reduction provides an
explicit solution for mapping (\ref{4dm}). If the ($n-1$)st iterate of the mapping (\ref{4dm})
is denoted $(x_n,x_{n+1},x_{n+2},x_{n+3})$, then
\begin{eqnarray*}
x_n&=&x_1
+ \lfloor\frac{n+2}{4}\rfloor(x_2-x_1)
+ \lfloor\frac{n+1}{4}\rfloor(x_3-x_2)
+ \lfloor\frac{n}{4}\rfloor(x_4-x_1) \\
&& + \lfloor\frac{n-1}{4}\rfloor \frac{(x_2-x_1)(x_4-x_3)}{x_3-x_2},
\end{eqnarray*}
where $\lfloor \ \rfloor$ denotes the floor function, sending $x$ to the largest integer below $x$.
This solution can be obtained similarly to the solution of the (3,1)-reduction of lattice pKdV given
in the appendix of \cite{grods}.

\subsubsection{The QD-algorithm}
As the QD-system (11) admits 1 (scaling) symmetry, all
mappings can be $1$-reduced. The reduced mapping has one integral
less, as only homogeneous combinations with scaling eigenvalue 0 are
invariant under scaling. For example, the mapping (\ref{sam}) with
new variables $y_i=x_i/x_4$, $i=1,2,3$, reduces to the 3-dimensional
mapping
\begin{equation} \label{3dm}
(y_1,y_2,y_3) \mapsto \left(\frac{y_3}{y_2}, y_1 + \frac{1-y_3}{y_2}, y_1\right),
\end{equation}
which has invariants
\[
I_1=\frac{y_1y_3}{y_2-y_3+1}, \qquad I_2=\frac{(y_2 - y_3) (y_1 - 1) (y_3 - 1)}{(y_2-y_3+1)^2}.
\]

This mapping has a lot of periodic points.
The first few are given in Table \ref{PO}, where $a,b$ are
free parameters. Also, one can show that the orbit of $(a,b,b)$
converges to the periodic orbit of $(1,ab,1)$, which has length 3.

\begin{center}
\begin{tabular}{l|l}
orbit length & periodic points \\
& \\
 \hline
& \\
1 & (0,-1,0),\ ($a,1,a$) \\
2 & (0,1,2),\ (2,-1,0),\ ($0,a,0$) \\
3 & ($1,a,1$),\ ($1,a,a$),\ ($a,1,1$) \\
4 & ($a,b-1,b$) \\
5 & ($a,b(b+1)/(a+b-1),b$) \\
\end{tabular}
\captionof{table}{\label{PO} Periodic points with orbit length smaller than six.}
\end{center}

Introducing variables $x=y_1,y=y_3$ on a level set of the first invariant $I_1=z$ yields the mapping
\[
\tau: (x,y) \mapsto (\frac{zy}{xy+zy-z},x).
\]
This map can be written as a composition of two involutions, namely $\tau=i_0 \circ i_1$, where
\[
i_0 : (x,y)\mapsto (y,x), \qquad i_1 : (x,y) \mapsto \left( \frac{f_1-f_2 x}{f_2-f_3 x}, y \right)
\]
with $f=Av\times Bv$, $v=(y^2,y,1)^t$, which has invariant
\[
\frac{w \cdot Av}{w \cdot Bv}=\frac{I_2}{z},
\]
where $w=(x^2,x,1)$. Thus, it is a special case of the 18-parameter QRT-family of planar maps \cite{QRT1,QRT2},
with
\[
A=\begin{pmatrix} 1 & -1 & 0 \\ -1 & 1-z & z \\ 0 & z & -z \end{pmatrix}, \qquad
B=\begin{pmatrix} 1 & 0 & 0 \\ 0 & 0 & 0 \\ 0 & 0 & 0 \end{pmatrix}.
\]

Reduction of order also works for correspondences. In
reduced coordinates $y_i=x_{i+1}/x_1$, $i=1,2,3$, the four dimensional mappings $\Pm_4,\Rm_4$ reduce to
\begin{equation} \label{3pm}
\Pm:\ (y_1,y_2,y_3)\mapsto \left( \frac{1}{y_3}, \frac{y_1}{y_3}, \frac{y_2}{y_3} \right)
\end{equation}
and
\begin{equation} \label{3rm}
\Rm:\ (y_1,y_2,y_3)\mapsto \left( y_3 \frac{(y_1+1)^2}{(y_3+y_2)^2}, y_2\frac{(y_1+1)^2}{(y_3+y_2)^2}, y_1\right),
\end{equation}
which admit the invariants
\[
\frac{y_2+y_1y_3}{(1+y_1+y_2+y_3)^2},\quad \frac{y_1y_2y_3}{(1+y_1+y_2+y_3)^4}.
\]

\subsection{Mappings with insufficiently many integrals}
We will next encounter reductions whose dimension is greater
than twice the number of functionally independent integrals
provided by the staircase method.
\subsubsection{The potential Korteweg-de Vries equation}
\label{skdv}

The matrices
\[
\left( \begin{array}{cc}
u & -k-u\t{u} \\
1 & -\t{u}
\end{array} \right), \quad
\left( \begin{array}{cc}
u & -\alpha - k-u\h{u} \\
1 & -\h{u}
\end{array} \right),
\]
form a Lax pair for the lattice pKdV equation
\begin{equation} \label{pkdv}
(u-\h{\t{u}})(\t{u}-\h{u})=\alpha.
\end{equation}

\subsubsection{($n-1,1$)-reduction}

\noindent
\parbox{40mm}{
\setlength{\unitlength}{8mm}
\begin{center}
\begin{picture}(3,1)(0,0)
\matrixput(0,0)(1,0){4}(0,1){2}{\circle{.1}}
\path(1,0)(1,1)(2,1)(2,0)(1,0)
\dottedline{.12}(-.5,0)(1,0)(1,1)(3.5,1)
\multiput(0,0)(1,0){2}{\circle*{.1}}
\multiput(1,1)(1,0){3}{\circle*{.1}}
\put(0,0.2){$x_{n-1}$}
\put(1.1,0.1){$x_n$}
\put(1,1.2){$x_1$}
\put(2,1.2){$x_2$}
\put(3,1.2){$x_3$}
\put(2.1,0.1){$\t{x_n}$}
\end{picture}
\captionof{figure}{\label{kdvlsc}}
\end{center}
} \hfill \parbox{75mm}{
We consider the initial value problem
\[
u_{(n-1)k+i-1,k}=x_{i},
\]
with $i=1,\ldots,n$ and $k\in\Z$, see Figure \ref{kdvlsc}.
We find, for all $k$, $u_{(n-1)k+n,k}=x_{1}+\alpha/(x_n-x_2).$
}

\smallskip
\noindent
Hence, the right-shift induces an $n$-dimensional mapping,
\begin{equation} \label{den}
 (x_1,x_2,\ldots,x_{n}) \mapsto \left(x_2,x_3,\ldots,x_{n},x_1+\frac{\alpha}{x_n-x_2}\right),
\end{equation}
which is volume-preserving when $n$ is odd and
anti-volume-preserving when $n$ is even.

We have verified up to $n=17$ that the trace of the monodromy matrix provides
$\lfloor (n-1)/2 \rfloor$ functionally independent integrals.
So, for the odd dimensional mappings we need one reduction, but for the even dimensional
mappings we need two. In fact, as we shall see, in the even case there exist three. How to
explain this? As we will see below, the KdV-equation has a Lie-point symmetry that does not
depend on the lattice variables. This symmetry gives rise to one reduction for both the odd
and the even dimensional mappings. Also, there are two symmetries that do depend on the lattice
variables. These yield two $2$-symmetries of the mapping if and only if its dimension is even,
giving us two more reductions.

Equation (\ref{pkdv}) has
the following symmetry $u\mapsto u+\epsilon$. This yields a symmetry
for the mapping (\ref{den}), whose infinitesimal generator is
\begin{equation} \label{fvf}
\sum_{i=1}^n \frac{\partial}{\partial x_{i}}.
\end{equation}
Let $y_i=x_i-x_{i+1}$ for $i=1,2,\ldots,n-1$. The functions $y_i$ are annihilated by the vector field (\ref{fvf}), they form a set of $n-1$ functionally independent invariants of the symmetry generated by this vector field. Taking the $y$ as new variables the mapping (\ref{den}) reduces to
\begin{equation} \label{pkdvmap}
(y_1,y_2,\ldots,y_{n-1}) \mapsto \left( y_2,y_3,\ldots,y_{n-1},
-\sum_{i=1}^{n-1} y_i + \frac{\alpha}{\sum_{i=2}^{n-1} y_i} \right).
\end{equation}
In addition, equation (\ref{kdv}) also has the following two symmetries
\[
u_{l,m} \mapsto u_{l,m} -(-1)^{l+m} \epsilon, \qquad u_{l,m}
\mapsto u_{l,m}\epsilon^{(-1)^{l+m}}.
\]
Suppose now that $n$ is even. Then the above symmetries of the P$\Delta$E (\ref{pkdv})
give rise to 2-symmetries of the mapping (\ref{pkdvmap}), with generators
\begin{equation} \label{stvf}
\sum_{i=1}^n (-1)^i \frac{\partial}{\partial x_{i}}, \qquad
\sum_{i=1}^n (-1)^i x_i \frac{\partial}{\partial x_{i}}.
\end{equation}
It can be verified that of $n-3$ functionally independent
joint invariants of the above three vector fields (\ref{fvf}, \ref{stvf})
are
\[
q_i=(x_i-x_{i+2})(x_{i+1}-x_{i+3}),\quad i=1,\ldots,n-3.
\]
We will take the $q_i$ as reduced variables and perform the reduction.
Let us define, with $k,m\in\N$,
\[
F^m_k:=\frac{x_{1+k}-x_{2m+3+k}}{x_{2m+1+k}-x_{2m+3+k}}.
\]
The $F^m_k$ satisfy the recurrence relation
\begin{equation} \label{rec}
(F^m_k-1)q_{2m+k}=q_{2m+k-1}F^{m-1}_k,
\end{equation}
with initial condition $F^0_k=1$. Therefore the $F^m_k$ can be expressed
in the $q_i$, with $i\leq n-3$, when $m<n/2-1$ and $k<n/2-2m+3$. We
have the following equalities
\begin{align*}
x_i&=x_{n-1}+\frac{q_{n-3}}{x_{n-2}-x_n}F^{(n-3-i)/2}_{i-1}, &i \text{ odd}, \\
x_i&=x_{n-2}+\frac{q_{n-4}}{q_{n-3}}(x_{n-2}-x_n)F^{(n-4-i)/2}_{i-1}, &i \text{ even},
\end{align*}
which is an inverse reduction. In terms of reduced variables the mapping is
\begin{align}
q_i &\mapsto q_{i+1},\ \ \ i\in\{1,2,\ldots,n-4\}, \notag \\
q_{n-3}%&=(x_{n-3}-x_{n-1})(x_{n-2}-x_n) \notag \\
&\mapsto %(x_{n-2}-x_{n})(x_{n-1}-x_1+\alpha(x_2-x_n)^{-1}) \notag\\ &=
- q_{n-3} F^{(n-4)/2}_0 + \alpha / F^{(n-4)/2}_1. \label{map}
\end{align}
An explicit expression for $F^m_k$ in terms of $q_i$ is\footnote{
The function $F^m_k$ is closely related to the function $Q$ given by equation (\ref{Q}),
we have $F^m_k-1=Q^{m-2}_{k+1}(q)(\prod_{j=1}^{k}q_{2j+m})^{-1}$.}
\begin{equation} \label{fmk}
F^m_k=\sum_{i=1}^{m+1} \prod_{j=2i}^{2m+1} q_{j-1+k}^{(-1)^j},
\end{equation}
since this expression solves the recurrence (\ref{rec}) with
$F^0_k=1$.
The mapping (\ref{map}) is anti-measure preserving with density $\prod_{i=1}^{(n-4)/2} q_{2i}$.
%The monodromy matrix is given by
%\[
%M(x_n,x_1)L(x_{n-1},x_n)\cdots L(x_2,x_3)L(x_1,x_2).
%\]
%We have checked up to $n=16$ that its trace provides us with $n/2-1$ functionally
%independent integrals for the mapping \ref{map}.

At $n=4$ the reduced mapping is $q_1 \mapsto \alpha - q_1$, which admits one
integral, $q_1(\alpha-q_1)$. The second iterate of this mapping equals the
identity. Note that this enables one to explicitly solve equation (\ref{den}) with $n=4$, cf. \cite{grods}.
What happened? Well, the joint invariant $q_1$ turns out
to be a 2-integral of the mapping. Let us define another set of
functions
\[
H^m_k := - x_{1+k} x_{2m+2+k} - \sum_{i=1}^{2m+1} (-1)^i x_{i+k}
x_{i+1+k}.
\]
They can be expressed in terms of the $q_i$, with $i\leq n-3$, if
$m<n/2$ and $k<n/2+4-2m$, using the recurrence
\begin{equation} \label{rec2}
H^m_k=H^{m-1}_k+q_{2m-1+k}F^{m-1}_k,
\end{equation}
with initial condition $H^0_k=0$. For all $n$ the $n$-dimensional
mapping $\delta_n$, see (\ref{den}), admits the 2-integral $H^{(n-2)/2}_0$. If the image of $x$ under (\ref{den}) is denoted $\t{x}$, then
\[
\widetilde{H^{(n-2)/2}_0}=\alpha-H^{(n-2)/2}_0.
\]
An explicit expression for $H^m_k$ in terms of the $q_i$ is
\[
H^m_k=\sum_{i=1}^m \sum_{j=1}^{m+1-i} \prod_{l=0}^{2i-2}
q_{2j+l-1+k}^{(-1)^l},
\]
since this solves the initial value problem (\ref{rec2}).
At $n=6$ the 3-reduced mapping is
\begin{equation} \label{n=6}
(q_1,q_2,q_3) \mapsto (q_2,q_3,-\frac{q_3(q_1+q_2)}{q_2}+\alpha\frac{q_3}{q_2+q_3}).
\end{equation}
The staircase method provides two integrals
\[
q_1q_3(q_2+q_3-\alpha+\frac{q_1(q_2+q_3)}{q_2}),\
(q_1+q_3+\frac{q_1q_3}{q_2})(q_1+q_3-\alpha+\frac{q_1q_3}{q_2}),
\]
of which the latter can be expressed in terms of the 2-integral
$H^2_0=q_1+q_3+q_1q_3/q_2$. We can take the 2-integral as a
variable. In terms of $q_1,q_2$ and $p=H^{(n-2)/2}_0$ mapping (\ref{n=6})
becomes
\[
(q_1,q_2,p)\mapsto (q_2,q_2(p-q_1)/(q_1+q_2),\alpha-p)
\]
which has integrals
\[
\frac{q_1q_2}{q_1+q_2}(p-q_1)(p+q_2-\alpha),p(\alpha-p).
\]
In general, with $n>2$ even, the 3-reduced mapping can be written as
\[
(q_1,q_2,\ldots,q_{n-4},p)\mapsto
(q_2,q_3,\ldots,q_{n-4},\frac{p-H^{(n-4)/2}_0}{F^{(n-4)/2}_0},\alpha-p),
\]
where we used (\ref{rec2}) to solve $p=H^{(n-2)/2}_0$ for $q_{n-3}$.

We note that all $\lfloor (n-1)/2 \rfloor$ functionally independent integrals
we have calculated for $n\leq17$ dimensional mappings survive these reductions.
Taking $n=2m+1$ odd, the reduced mapping is $2m$ dimensional and has $m$ integrals.
With $n=2m+2$, the reduced mapping is $2m-1$ dimensional and has $m$ integrals.

\subsection{The Boussinesq system} \label{sB}

\def\22{34}

\noindent
\parbox{25mm}{
\setlength{\unitlength}{10mm}
\begin{center}

\begin{picture}(1.8,6)(0,0)
\matrixput(.05,0)(1,0){2}(0,1){7}{\circle{.1}}
\matrixput(-.05,0)(1,0){2}(0,1){7}{\circle{.1}}
\matrixput(0,-.05)(1,0){2}(0,1){7}{\circle{.1}}
\multiput(.95,0)(0,1){7}{\circle*{.1}}
\multiput(-.05,1)(0,1){3}{\circle*{.1}}
\put(-.05,6){\circle*{.1}}
\put(.05,6){\circle*{.1}}
\multiput(.05,1)(1,-1){2}{\circle*{.1}}
\multiput(1.05,3)(0,1){2}{\circle*{.1}}
\put(0,1.95){\circle*{.1}}
\multiput(1,4.95)(0,1){2}{\circle*{.1}}

\path(1,0)(1,1)(0,1)(1,0)
\path(0,2)(1,2)(1,3)(0,3)(0,2)
\path(1,4)(1,5)
\path(0,6)(1,6)
\multiput(.7,0.1)(0,1){7}{$u$}
\multiput(-.3,1.1)(0,1){3}{$u$}
\put(-.3,6.1){$u$}
\multiput(.1,1.1)(0,5){2}{$v$}
\put(.1,2.1){$w$}
\put(1.1,.1){$v$}
\multiput(1.1,3.1)(0,1){2}{$v$}
\multiput(1.1,5.1)(0,1){2}{$w$}
\put(1.8,6){$(a)$}
\put(1.8,4.5){$(b)$}
\put(1.8,2.5){$(c)$}
\put(1.8,.5){$(d)$}
\end{picture}
\captionof{figure}{\label{bsq}}
\end{center}
}
\hfill \parbox{90mm}{
The Boussinesq system \cite{NPCQ,TN}
\begin{align}
\t{w} &= u \t{u} - v \tag{{\22}a} \\
\h{w} &= u \h{u} - v \tag{{\22}b} \\
w &= u \h{\t{u}} - \h{\t{v}} +
\frac{\gamma}{\t{u} -\h{u}}. \tag{{\22}c} \addtocounter{equation}{1}
\end{align}
is defined on the square as depicted in Figure \ref{bsq}(a,b,c).
In Figure \ref{bsq}(d) we have depicted the consequence of equations
({\22}a,{\22}b),
\begin{equation}
\h{u}\h{\t{u}}-\h{v}=\t{u}\h{\t{u}}-\t{v} \tag{{\22}d}.
\end{equation}
From the Boussinesq system one can eliminate the variables $v$ and
$w$ using the identity
\[
(\t{\h{v}} + w)^{\h{}} -
(\t{\h{v}} + w)^{\t{}} =
(\h{v}-\t{v})^{\t{\h{}}} +
(\h{w}-\t{w})
\]
to get a 9-point scalar equation on a $2\times2$ square, called the
Boussinesq equation, cf. \cite[equation 1.3]{NPCQ}.}

\smallskip
\noindent
We denote $\u=(u,v,w)$. A Lax-pair for the system ({\22}) is given by, cf. \cite{TN},
\begin{equation}\label{Lp}
\begin{split}
L_\mathbf{u}=L(u, w, \t{u}, \t{v})
&=\left(\begin{array}{ccc} -\t{u} & 1 & 0 \\
-\t{v} & 0 & 1 \\ \t{u}w-\t{v}u - k & -w & u
\end{array}\right), \\
M_\mathbf{u}=M(u, w, \h{u}, \h{v})
&=\left(\begin{array}{ccc}
-\h{u} & 1 & 0 \\ -\h{v} & 0 & 1 \\
\h{u}w-\h{v}u + \gamma - k & -w & u \end{array}\right).
\end{split}
\end{equation}
Invariants for traveling wave reductions of the system can be obtained
by expanding traces of powers of the monodromy matrix.
Since $\cL$ is a $3\times 3$ matrix  a full set of functionally independent
integrals can be obtained from
$k$-expansions of the coefficients in
\begin{align}
Det(\lambda I - \cL) = &\lambda^3 - \lambda^2 \tr(\cL) + \lambda \frac{\tr(\cL)^2-\tr(\cL^2)}{2}\notag \\
&- \frac{\tr(\cL)^3- 3 \tr(\cL)\tr(\cL^2) +2 \tr(\cL^3)}{6}. \label{lamex}
\end{align}
cf. section \ref{sm}, in particular equation (\ref{vie}). However, due to the fact that both Lax-matrices
have a constant determinant, it suffices to consider $\tr(\cL)$ and $\tr^2(\cL)-\tr(\cL^2)$.

The following proposition tells us how to pose initial value problems
for the Boussinesq system. The proof uses a different technique than
the one used in \cite{K}, which is possible due to the fact that for
the Boussinesq system initial values can be given on staircases.
However, the staircases are not necessarily {\em standard} staircases, which
they would be in the framework of \cite{K} (at least for equations defined
on the square, such as the Boussinesq system).

We call a staircase {\bf ascending}, if it goes from the lower left to the upper right,
that is, if it is a sequence of neighboring lattice sites with $l$ and $m$ nondecreasing. And we
call a staircase {\bf descending} if it goes from the upper left to
the lower right, that is, if it is a sequence of neighboring lattice
sites with $l$ and $-m$ nondecreasing.
\begin{proposition} \label{Pbsq}
The following initial values problems for the Boussinesq system are
well-posed.
\begin{itemize}
\item At every point on an ascending staircase take the components
$u,v$ of the vector $\mathbf{u}$ as initial values.
\item On a descending staircase take $u,v,w$ at the lower
left corners, $v$ at the upper right corners, and $u,v$ at
the other points as initial values.
\end{itemize}
\end{proposition}
\proof The proof consist of two parts. Firstly, we show that the values
at all points of the staircase can be obtained from the initial
values. Secondly we show that any of the four vectors
$\mathbf{u},\t{\mathbf{u}},\h{\mathbf{u}},\h{\t{\mathbf{u}}}$,
can be determined from the other three.
\begin{itemize}
\item[$i)$]
For ascending staircases the first part is easy. Going along the
staircase from the lower-left to the upper-right at each horizontal
step the component $w$ is obtained using equation (${\22}a$), whereas at
the vertical steps equation (${\22}b$) can be used. For a descending
staircase we can do a similar thing, except at the upper-right
corners.
Equation (${\22}d$) can be solved for $u_{l+1,m+1}$ and used to get
the $u$-components at the upper-right corners. Once $u$ has been
calculated $w$ can be calculated in two ways, using either (${\22}a$) or
({\22}b), leading to the same result.
\item[$ii)$] The values of $\h{\mathbf{u}}$ can be obtained as follows.
First calculate $\h{u}$
from (${\22}c$). Then $\h{w}$ can be obtained from (${\22}b$) and
$\h{v}$ from the up-shifted consequence of (${\22}a$). We can obtain
$\t{\mathbf{u}}$ in a similar way. This follows from the fact
that interchanging the left-shift with the up-shift and
$\gamma\mapsto -\gamma$ is a (discrete) symmetry of the system.
Finally, to obtain $\h{\t{\mathbf{u}}}$ one uses the
consequence (${\22}d$) to calculate $\h{\t{u}}$, after which
$\h{\t{w}}$ is found using a shifted version of either (${\22}a$)
or (${\22}b$), and $\h{\t{v}}$ is calculated using (${\22}c$). Finally,
due to the discrete symmetry which interchanges the up-shift with
the down-shift and the left-shift with the right-shift, together
with $u\leftrightarrow w$ and $\gamma\mapsto -\gamma$, it follows
that $\mathbf{u}$ can be obtained from given values at the other
sites.
\end{itemize}
\qed
\noindent
It follows that, with $s=(s_1,s_2)\in\Z \times \Z$ such that $s_1s_2\neq0$,
the dimension of an $\s$-periodic reduction is $2(\abs{s_1}+\abs{s_2})$.

\subsubsection*{(n-1,1)-reduction}
\noindent
\parbox{55mm}{
\setlength{\unitlength}{14mm}
\begin{center}
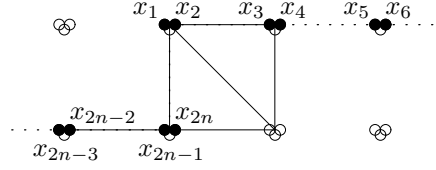

\begin{picture}(3,1.5)(-.3,-.3)
\matrixput(-.05,0)(1,0){4}(0,1){2}{\circle{.1}}
\matrixput(0.05,0)(1,0){4}(0,1){2}{\circle{.1}}
\matrixput(0,-.05)(1,0){4}(0,1){2}{\circle{.1}}
\dottedline{.12}(-.5,0)(0,0)(1,0)(1,1)(3,1)(3.5,1)
\multiput(-.05,0)(1,0){2}{\circle*{.1}}
\multiput(.95,1)(1,0){3}{\circle*{.1}}
\multiput(0.05,0)(1,0){2}{\circle*{.1}}
\multiput(1.05,1)(1,0){3}{\circle*{.1}}
\put(-0.3,-0.25){$x_{2n-3}$}
\put(0.05,0.1){$x_{2n-2}$}
\put(0.7,-0.25){$x_{2n-1}$}
\put(1.05,0.1){$x_{2n}$}
\put(.65,1.1){$x_1$}
\put(1.05,1.1){$x_2$}
\put(1.65,1.1){$x_3$}
\put(2.05,1.1){$x_4$}
\put(2.65,1.1){$x_5$}
\put(3.05,1.1){$x_6$}
\path(0,0)(2,0)(2,1)(1,1)(1,0)
\path(1,1)(2,0)
\end{picture}
\captionof{figure}{$2n$ initial values \label{Bsas}}
\end{center}
} \hfill \parbox{55mm}{
We take $s_1=n-1$ positive (but $n\neq 3$), $s_2=1$, and we consider the
following initial value problem, with $i=0,1,\ldots,n-1$ and $l\in\Z$.
\begin{align*}
u_{(n-1)l+i,l}=x_{2i+1},\\
v_{(n-1)l+i,l}=x_{2i+2},
\end{align*}
where the index on $x$ is taken modulo $2n$, see Figure \ref{Bsas}.
}

\smallskip
\noindent
Using the equations (${\22}a$), (${\22}c$), and (${\22}d$), in that order, we calculate, $w_{(n-1)l,l-1}=x_{2n-3}x_{2n-1}-x_{2n-2}$,
$u_{(n-1)l+1,l-1}=x_1 +\gamma P$, and $v_{(n-1)l+1,lk-1}=x_2 + x_3 \gamma P$, where
\begin{equation} \label{P}
P=\frac{1}{x_{2n-1}(x_{2n-3}-x_3)-x_{2n-2}+x_4}.
\end{equation}
The right-shift induces the $2n$-dimensional mapping $\map_{2n}:$,
\begin{align}
x_i &\mapsto x_{i+2}, \quad i\in\{1,2,\ldots,2n-2\}, \notag \\
x_{2n-1} &\mapsto x_1+\gamma P, \label{2nb} \\
x_{2n}&\mapsto x_2+ \gamma x_3 P. \notag
\end{align}
The monodromy matrix is
\[
\cL=M(x_{2n-1},w_n,x_1,x_2)L(x_{2n-3},w_{n-1},x_{2n-1},x_{2n})\cdots
L(x_1,w_1,x_3,x_4),
\]
where $w_1=x_1x_{2n-1}-x_{2n}$ and $w_{i+1}=x_{2i-1}x_{2i+1}-x_{2i}$, $i=1,2,\ldots,n-1$.

\smallskip
\noindent
\parbox{55mm}{
\setlength{\unitlength}{14mm}
\begin{center}
\begin{tabular}{c|c|c|c|c|c|c|c|c}
$n$ & 2 & 3 & 4 & 5 & 6 & 7 & 8 & 9\\
% & & & & & & & & \\
 \hline
% & & & & & & & & \\
\# & 1 & 0 & 3 & 4 & 3 & 6 & 7 & 6\\
\end{tabular}
\captionof{table}{\label{nofii} Number of functionally independent
integrals of $\map_{2n}$.}
\end{center}
} \hfill \parbox{55mm}{
The number of functionally independent integrals we have obtained for reductions
with period $\s=(n-1,1)$ is given in table \ref{nofii}. From this table it seems
we need to $d$-reduce mapping $\map_{2n}$ by $d=2$ dimensions, or, if $3$ divides $n$,
by $d=6$ dimensions.
}

\smallskip
\noindent
The mapping $\map_{2n}$ has two symmetries, generated by
\[
v_1=\sum_{i=1}^n \frac{\partial}{\partial x_{2i}},\qquad
v_2=\sum_{i=1}^n \frac{\partial}{\partial x_{2i-1}} +
x_{2i-1}\frac{\partial}{\partial x_{2i}}.
\]
The easiest way to check that these vector fields are generators of symmetries indeed is using
the Jacobian, we have $Jv=\map_{2n}(v)$ when
$v=(0,1,0,1,\ldots,0,1)$ or $v=(1,x_1,1,x_3,\ldots,1,x_{2n-1})$, where the
Jacobian matrix of $\map_{2n}$ is given by
\[
J= \left(
\begin{array}{cc}
0 &I_{2n-2} \\
I_{2}& \gamma P^2 H
\end{array}
\right)
\]
where $I_k$ is the $k\times k$ identity matrix and $H$ is the
$2\times (2n-2)$ matrix
\[
\left(
\begin{array}{ccccccccc}
x_{2n-1} & -1 & 0 & \cdots & 0 & -x_{2n-1} & 1 & (x_3-x_{2n}-3) & 0 \\
1/P + x_3x_{2n-1} & -x_3 & 0 & \cdots & 0 & -x_3x_{2n-1} & x_3 &
x_3(x_3-x_{2n}-3) & 0
\end{array}
\right) .
\]
The two symmetries of the mapping $\map_{2n}$ correspond to the following
symmetries of the original lattice system (\22):
\begin{equation} \label{sym}
(u,v,w)\mapsto (u,v+\epsilon,w-\epsilon),\qquad (u,v,w)\mapsto
(u+\epsilon,v+\epsilon u,w+\epsilon u).
\end{equation}
It can be verified that the functions
\[
y_i=x_{2i-1}-x_{2i+1},y_{n+i-1}=x_{2i}-x_{2i+2}+x_{2i-1}(x_{2i+1}-x_{2i-1}),
\]
with $i=1,\ldots, n-1$, are joint invariants of the these symmetries and functionally independent. In the reduced variables $y$, we get a $2(n-1)$-dimensional volume preserving mapping
\begin{align*}
y_i &\mapsto y_{i+1},\quad i\in\{1,2,\ldots,2n-2\}, i\neq n-1, \\
y_{n-1} &\mapsto -\sum_{i=1}^{n-1} y_i - \gamma Q,\\
y_{2n-2}&\mapsto -(\sum_{i=0}^{n-2}
y_{n+i}+y_{i+1}\sum_{j=i+1}^{n-1}y_j ) - \gamma Q \sum_{i=2}^{n-1}
y_i,
\end{align*}
where $Q=\sum_{i=1}^{n-3} (y_{n+i} + y_{i+1}\sum_{j=i+1}^{n-1}y_j)$.

When $n=2$ the reduced mapping is, in terms of $X=-y_1,Y=-y_2$
\[
(X,Y) \mapsto (-X+\gamma(Y-X^2),-Y+X^2)
\]
which carries the invariant $Y(X^2 - Y)+\gamma X$, cf.
\cite[equation 5.31]{NPCQ}. Note, in \cite{NPCQ}
the case $s_1=s_2$ was studied, in particular the involutivity of the integrals was established
in any dimension. The authors defer the actual counting of independent integrals to a future study.
However, they also state that the investigation of lower-dimensional examples (with $s_1=s_2$) indicate
a sufficient number of invariants are functionally independent.

We have verified that all integrals we found, see Table \ref{nofii},
survive the 2-reduction. Hence, in those cases, except when $3$ divides $n$, the staircase method
provides enough integrals for integrability. Next we will show that if $3$ divides
$n$, but $n\neq3$, we can further reduce the mapping by four dimension.

The Boussinesq system ({\22}) has some additional symmetries, which
depend on the lattice variables ($\mathbf{u}=\mathbf{u}_{l,m}$),
\begin{align}
%(u,v,w)_{l,m} &\mapsto (u,v+\epsilon \zeta^{l+m+1},w-\epsilon\zeta^{l+m})_{l,m}, \notag \\
%(u,v,w)_{l,m} &\mapsto
%(u+\epsilon\zeta^{l+m},v+\epsilon\zeta^{l+m+1} u,w+\epsilon
%\zeta^{l+m-1} u)_{l,m}, \label{sdlv}
(u,v,w)&\mapsto (u,v+\epsilon \zeta^{l+m+1},w-\epsilon\zeta^{l+m}), \notag \\
(u,v,w) &\mapsto
(u+\epsilon\zeta^{l+m},v+\epsilon\zeta^{l+m+1} u,w+\epsilon
\zeta^{l+m-1} u), \label{sdlv}
\end{align}
where $\zeta$ is a primitive third root of unity, that is,
$\zeta^2+\zeta+1=0$. The generators of the corresponding
transformations acting on the initial values (\ref{Bsas}) are
\[%\begin{align*}
v_3=\sum_{i=1}^n \zeta^i \frac{\partial}{\partial x_{2i}}, \quad
v_4=\sum_{i=1}^n \left( \zeta^{i-1} \frac{\partial}{\partial x_{2i-1}} +
\zeta^i x_{2i-1}\frac{\partial}{\partial x_{2i}} \right),
\]%\end{align*}
and, taking the conjugate root $\zeta^2$,
\[%\begin{align*}
v_5=\sum_{i=1}^n \zeta^{2i} \frac{\partial}{\partial x_{2i}}, \quad
v_6=\sum_{i=1}^n \left( \zeta^{2i+1} \frac{\partial}{\partial x_{2i-1}} +
\zeta^{2i} x_{2i-1}\frac{\partial}{\partial x_{2i}} \right).
\]%\end{align*}
Now assume that 3 divides $n$. We construct real vector fields by taking the following linear
combinations, with $i=1,2,3$,
\begin{align*}
w_i=(v_1+\zeta^{2i}v_3+\zeta^i v_5)/3
&=\sum_{j=0}^{n/3-1} \frac{\partial}{\partial x_{6j+2i}} \\
w_{3+i}=(v_2+\zeta^{2i}v_4+\zeta^i v_6)/3
&=\sum_{j=0}^{n/3-1} x_{6j+2i-1} \frac{\partial}{\partial x_{6j+2i}}
+\frac{\partial}{\partial x_{6j+2i+1}},
\end{align*}
where $x_{2n+1}=x_1$. These vector fields are 3-symmetries of
the mapping $\map_{2n}$. Let $J^3$ be the Jacobian matrix of $\map_{2n}^3$. Note, the vector
fields $w_i$, $i\neq 1,4$ can be obtained from
$w_i=J^3\map_{2n}(w_{i-1})$. According to \cite[proposition 1]{HBQC} it
suffices to verify that $w_1$ and $w_4$ are 3-symmetries. Also note
that $J^3\map_{2n}v_i=v_i$ for $i\in\{1,\ldots,6\}$.

The following polynomials form a complete set of joint invariants of
the vector fields $w_1,w_2,\ldots,w_6$:
\begin{align*}
z_i&=x_{2i-1}-x_{2i+5},\quad i\in\{1,\ldots,n-3\}, \\
z_{n-3+i}&=x_{2i+1}(x_{2i-1}-x_{2i+5})-x_{2i}+x_{2i+6},\quad i\in\{1,2,3\}, \\
z_{n+i}&=(x_{2i-1}-x_{2i+5})(x_{2i+6}-x_{2i+12}) \\
&\ \ \ -(x_{2i}-x_{2i+6})(x_{2i+5}-x_{2i+11}),\quad i\in\{1,2,\ldots,n-6\}.
\end{align*}
Another joint invariant is given by $P$, see (\ref{P}), which
therefore should be expressible in terms of the $z_i$. However, we haven't
found a general formula for $P(z)$. In terms of the $z$-variables the mapping
$\map_{12}$ reduces to
\begin{align*}
z_i &\mapsto z_{i+1},\quad i\in\{1,2,4,5\} \\
z_{3} &\mapsto -z_1+\frac{\gamma}{z_5-z_2z_3}, \\
z_{6} &\mapsto -z_4+z_1z_2-\gamma \frac{z_2}{z_5-z_2z_3},
\end{align*}
and $\map_{2n}$, with 3 divides $n>4$,
\begin{align*}
z_i &\mapsto z_{i+1},\quad i\in\{1,2,\ldots,2n-7\}, i\neq n-3, i\neq n \\
z_{n-3} &\mapsto -\sum_{i=1}^{n/3-1} z_{3i-2} - \gamma P, \\
z_{n} &\mapsto -\frac{z_{n+1}+z_4(z_1z_2-z_{n-2})}{z_1}, \\
z_{2n-6} &\mapsto A-\gamma B P,
\end{align*}
where $A=(x_{2n-11}-x_{2n-5})(x_{2n-4}-x_{2})-(x_{2n-10}-x_{2n-4})(x_{2n-5}-x_{1})$, and
$B=x_{3}(x_{2n-11}-x_{2n-5})-x_{2n-10}+x_{2n-4}$
are also joint invariants of the six vector fields $w_i$.
All the integrals we found survive the above reduction. Thus,
as one can see in Table \ref{nofii}, also when $3$ divides $n$ the staircase
method provides a sufficient number of integrals for the
6-reduced map to be integrable.

We found two functions which are 3-integrals, that is,
integrals of $\map^3_{2n}$. These are
\begin{align*}
i_1&=\sum_{j=1}^{n/3} (x_{6j+1}-x_{6j-5})(x_{6j-2}-x_{6j-3}x_{6j-1})
+ x_{6j-1}(x_{6j+2}-x_{6j-4}), \\
i_2&=\sum_{j=1}^{n/3}
x_{6j-1}(x_{6j-5}x_{6j-3}-x_{6j-4}-x_{6j+1}x_{6j+3}+x_{6j+2}) +
x_{6j}(x_{6j+3}-x_{6j-3}).
\end{align*}
We have the following action of $\map_{2n}$ on the 3-integrals
\begin{equation} \label{2d}
(i_1,i_2) \mapsto (i_2-i_1 + \gamma, -i_1 + \gamma ),
\end{equation}
whose third power is the identity. Note that by applying the map to
one of the 3-integrals gives us the other one but no third
functionally independent 3-integral can be obtained in this way. The
3-integrals admit the vector fields $w_i$ as symmetries and two of
the $n-3$ functionally independent integrals found by the staircase
method can be written in terms of them. For $n=6$ we have
\[
i_1=-z_1z_2z_3+z_1z_5+z_3z_4,\quad i_2=z_2z_6+z_3z_4.
\]
and two of the three functionally independent integrals found by the
staircase method are given by
\[
i_2^2+i_1(i_1-i_2-\gamma), i_1(\gamma-i_2)(\gamma-i_1+i_2)
\]
which are both integrals of the 2-dimensional map (\ref{2d}).
An extra advantage of working with expansion (\ref{lamex}) instead of
traces of powers of $\cL$ is that the third functionally
independent integral factorizes nicely as
\[
z_4z_5z_6(z_1z_3+z_6)(z_4z_5-z_4z_2z_3-z_2z_1z_5+z_2^2z_1z_3+\alpha z_2).
\]

\subsubsection*{(n-1,-1)-reduction}
\noindent
\parbox{60mm}{
\setlength{\unitlength}{14mm}
\begin{center}
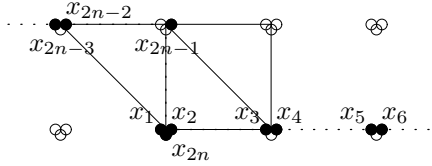

\begin{picture}(3,1.5)(0,-.5)
\matrixput(-.05,0)(1,0){4}(0,1){2}{\circle{.1}}
\matrixput(0.05,0)(1,0){4}(0,1){2}{\circle{.1}}
\matrixput(0,-.05)(1,0){4}(0,1){2}{\circle{.1}}
\dottedline{.12}(-.5,1)(0,1)(1,1)(1,0)(3,0)(3.5,0)
\put(-.05,1){\circle*{.1}}
\multiput(0.05,1)(1,0){2}{\circle*{.1}}
\multiput(.95,0)(1,0){3}{\circle*{.1}}
\multiput(1.05,0)(1,0){3}{\circle*{.1}}
\put(1,-0.05){\circle*{.1}}
\put(-0.3,.75){$x_{2n-3}$}
\put(0.05,1.1){$x_{2n-2}$}
\put(0.7,.75){$x_{2n-1}$}
\put(1.05,-.25){$x_{2n}$}
\put(.65,0.1){$x_1$}
\put(1.05,0.1){$x_2$}
\put(1.65,0.1){$x_3$}
\put(2.05,0.1){$x_4$}
\put(2.65,0.1){$x_5$}
\put(3.05,0.1){$x_6$}
\path(0,1)(2,1)(2,0)(1,0)(0,1)
\path(1,0)(1,1)(2,0)
\end{picture}
\captionof{figure}{Periodic initial value problem for the Boussinesq system \label{Bsds}}
\end{center}
} \hfill \parbox{60mm}{
We take $s_1=n-1$ positive ($n\neq 2$), $s_2=-1$, and we consider the following initial
value problem, with $i=0,1,\ldots,n-2$, $k\in\Z$,
\begin{align*}
u_{(n-1)k+i,-k}&=x_{2i+1},\\
v_{(n-1)k+i,-k}&=x_{2i+2},\\
v_{(n-1)(k+1),-k}&=x_{2n-1},\\
w_{(n-1)k,-k}&=x_{2n},
\end{align*}
see Figure \ref{Bsds}.}

\smallskip
\noindent
We right-shift the initial values $x_{2n-3}$ and $x_{2n}$ using equations (${\22}d$) and (${\22}a$), respectively. Then we right-shift $x_{2n-3}$ a second time and are able to determine the image of $x_{2n-1}$, using equation (${\22}c$).
Thus, we find the $2n$-dimensional mapping $\zeta_{2n}$:
\begin{align*}
x_i & \mapsto x_{i+2},\quad i\in\{1,2,\ldots,2n-3\}, \\
x_{2n-3}&\mapsto \frac{x_{2n-2}-x_2}{x_{2n-3}-x_1}, \\
x_{2n-2}&\mapsto x_{2n-1}, \\
x_{2n-1}&\mapsto -x_{2n} + \frac{(x_{2n-3}-x_1)(x_1(x_{2n-1}-x_4)
-\gamma)}{x_{2n-2}-x_2-x_3(x_{2n-3}-x_1)},\\
x_{2n} &\mapsto -x_2 + x_1 x_3,
\end{align*}
which is measure preserving with density $x_{2n-3}-x_1$ ($n\neq 2$).
The mapping $\zeta_{2n}$ admits the symmetries
\[
\frac{\partial}{\partial x_{2n-1}} - \frac{\partial}{\partial
x_{2i}} + \sum_{i=1}^{n-1} \frac{\partial}{\partial x_{2i}}
\]
and
\[
x_1\frac{\partial}{\partial x_{2n}} +
\zeta_{2n}(x_{2n-3})\frac{\partial}{\partial x_{2n-1}} + \sum_{i=1}^{n-1}
(\frac{\partial}{\partial x_{2i-1}} + x_{2i-1}\frac{\partial}{\partial x_{2i}}).
\]
Hence, it can be reduced to a ($2n-2$)-dimensional mapping. We have verified up to $n=7$, that
the number of functionally independent integrals, obtained by $k$-expansion
of the coefficients in (\ref{lamex}), with
\begin{align*}
\cL=M^{-1}(x_1,x_{2n},\zeta_n(x_{2n-3}),x_{2n-1})L(x_{2n-3},w_{n-1},\zeta_n(x_{2n-3}),x_{2n-1})\\
\cdot \prod_{i=2}^{\curvearrowleft \atop {n-1}}
L(x_{2i-3},w_{i-1},x_{2i-1},x_{2i}),
\end{align*}
is $n-1$, except when $3$ divides $n+1$, where the number is
$n-3$. Also we verified that all these integrals admit the above vector fields as their
symmetries.  If $3$ divides $n+1$ the symmetries (\ref{sdlv}) yield 3-symmetries of the mapping $\zeta_{2n}$.
As in the previous example we take linear combinations to get
\[
w_1=\sum_{j=0}^{(n-2)/3} \frac{\partial}{\partial x_{6j+2}}, \
w_4=(\sum_{j=0}^{(n-5)/3-1} x_{6j+1} \frac{\partial}{\partial
x_{6j+2}} +\frac{\partial}{\partial x_{6j+3}}) + x_{2n-3}
\frac{\partial}{\partial x_{2n-2}},
\]
%\begin{align*}
%w_1&=\sum_{j=0}^{(n-2)/3} \frac{\partial}{\partial x_{6j+2}}, \\
%w_4&=\left(\sum_{j=0}^{(n-5)/3-1} x_{6j+1} \frac{\partial}{\partial
%x_{6j+2}} +\frac{\partial}{\partial x_{6j+3}}\right) + x_{2n-3}
%\frac{\partial}{\partial x_{2n-2}},
%\end{align*}
together with $w_i=J^3 \zeta_{2n}(w_{i-1})$, $i\neq 1,4$. Taking $n=5$ we
obtain the following $4$-dimensional $6$-reduced mapping, in
terms of $ y_1=x_7-x_1, y_2=x_6+x_{10}-x_5x_1,
y_3=x_8-x_2+x_3(-x_7+x_1),
y_4=(-x_7+x_1)(-x_9+x_4-x_5x_3)+x_5(-x_8+x_2) $:
%\begin{align*}
%y_1 &\mapsto y_3/y_1, \\
%y_2 &\mapsto y_3, \\
%y_3 &\mapsto y_4/y_1, \\
%y_4 &\mapsto -y_2y_3/y_1+y_4 + \gamma.
%\end{align*}
\[
(y_1,y_2,y_3,y_4) \mapsto \left(\frac{y_3}{y_1}, y_3, \frac{y_4}{y_1},y_4 + \gamma-\frac{y_2y_3}{y_1}\right).
\]
This mapping is measure preserving with density  $y_1^2$. It has two 3-integrals, $i_1=y_2y_3/y_1$ and $i_2=y_4$,
which satisfy
\[
(i_1,i_2) \mapsto (i_2, -i_1-i_2-\gamma)
,
\]
whose third power is the identity. The two functionally independent
invariants found by the staircase method can be expressed in terms
of the 3-integrals as
\[
i_1^2+i_2^2+i_1i_2+\gamma (i_1+i_2), \quad i_1i_2(i_1+i_2+\gamma).
\]

\subsubsection{The potential Korteweg-De Vries equation, ($3,0$)-reduction} \label{pkdvh}
Consider initial values for the pKdV equation (\ref{pkdv}):
$u_{l,0}=x_{k}$ with $k\equiv l$ mod 3, $k\in\{1,2,3\}$, as in Figure \ref{kdvnlsc}.

\smallskip
\noindent
\parbox{55mm}{
\setlength{\unitlength}{10mm}
\begin{center}
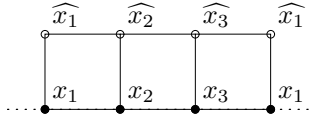

\begin{picture}(3,1.5)(0,0)
\matrixput(0,0)(1,0){4}(0,1){2}{\circle{.1}}
\thinlines
\path(2,1)(0,1)(0,0)(1,0)(1,1)
\path(1,0)(3,0)(3,1)(2,1)(2,0)
\dottedline{.12}(-.5,0)(3.5,0)
\multiput(0,0)(1,0){4}{\circle*{.1}}
\multiput(0.1,0.15)(3,0){2}{$x_1$}
\multiput(1.1,0.15)(3,0){1}{$x_2$}
\multiput(2.1,0.15)(3,0){1}{$x_3$}
\multiput(0.1,1.15)(3,0){2}{$\h{x_1}$}
\multiput(1.1,1.15)(3,0){1}{$\h{x_2}$}
\multiput(2.1,1.15)(3,0){1}{$\h{x_3}$}
\end{picture}
\captionof{figure}{\label{kdvnlsc} Initial values on a horizontal staircase with periodicity $u_{l,m}=u_{l+3,m}$.}
\end{center}
} \quad \hfill \parbox{60mm}{
Updating these by shifting them upwards, and imposing the image to be periodic, we have to solve the system
\begin{align}
(x_1-\h{x_2})(x_2-\h{x_1})&=\alpha \notag \\
(x_2-\h{x_3})(x_3-\h{x_2})&=\alpha \label{ste}\\
(x_3-\h{x_1})(x_1-\h{x_3})&=\alpha \notag
\end{align}
for the image points $\h{x_1},\h{x_2},\h{x_3}$.
}

\smallskip
\noindent
In terms of reduced variables $q_1=x_1-x_2$, $q_2=x_2-x_3$ the volume
preserving two-valued correspondence is $\mapt_{\pm}$:

\begin{align*}
q_1&\mapsto
q_1\frac{\alpha(\alpha+q_2(q_1+q_2))+q_1^2q_2(q_1+q_2)/2+(q_1/2+q_2)R}{(\alpha+q_1q_2)(\alpha-q_1(q_1+q_2))} \\
q_2&\mapsto
q_2\frac{\alpha(\alpha+q_1(q_1+q_2))+q_1q_2^2(q_1+q_2)/2-(q_1+q_2/2)R}{(\alpha+q_1q_2)(\alpha-q_2(q_1+q_2))},
\end{align*}
where
\[
R=\pm\sqrt{4\alpha^3+(q_1q_2(q_1+q_2))^2}.
\]
This correspondence admits the integral $q_1q_2(q_1+q_2)$, which can be obtained by taking the
trace of the monodromy matrix.

\section{The staircase method on quad-graphs}
Recently, in \cite{AV}, a geometric criterion was given for the well-posedness of initial
value problems on quad-graphs. In this section we show that for 'regular' quad-graphs,
those that permit periodic solutions, the staircase method can be applied. We use
equation $H3_{\delta=0}$, from \cite{ABS1},
\begin{equation} \label{H3}
Q_{pq}(a,b,c,d):=p(ab+cd)-q(ac+bd)=0,
\end{equation}
on two different quad-graphs, cf. Figure 9d and 9e in \cite{AV}.

We start with a brief introduction to the idea of a quad-graph. For a more thorough treatment and
references to the literature we refer to \cite{AV}. A quad-graph is a planar graph with quadrilateral faces.
Fields are assigned to the vertices and parameters to its edges. In the class considered in \cite{AV} opposite
edges carry the same parameters and the (multi-linear) equation is supposed to have $D_4$-symmetry (so that
the equation can be defined on each face independently of its position in the quad-graph).
Due to the first property there are sequences of adjacent quadrilaterals on which the value of the
parameter is constant. These are called {\it characteristics}. The main result in \cite{AV} states that
an initial value problem $P$ is well-posed if and only if each characteristic intersects $P$ in exactly
one edge.

\noindent
\parbox{55mm}{
\begin{center}
\setlength{\unitlength}{8mm}
\begin{picture}(4,2.5)(0,-0.5)
\put(2,1){\circle{.15}}
\multiput(1,2)(2,0){2}{\circle{.15}}
\multiput(0,1)(4,0){2}{\circle*{.15}}
\multiput(1,0)(2,0){2}{\circle*{.15}}
\thinlines 
\path(2,1)(1,0)(0,1)(1,2)(2,1)(4,1)(3,0)(1,0)
\path(1,2)(3,2)(4,1)
\dottedline{.14}(.5,.5)(1.5,1.5)(3.5,1.5)
\dottedline{.14}(.5,1.5)(1.5,.5)(3.5,.5)
\dottedline{.14}(2,0)(3,1)(2,2)
\put(-.3,.7){$a$}
\put(.7,-.3){$b$}
\put(3.1,-.3){$c$}
\put(4.1,.7){$d$}
\put(2.1,.7){$e$}
\put(.7,2.1){$f$}
\put(3.1,2.2){$g$}
\thicklines
\path(0,1)(1,0)(3,0)(4,1)
\end{picture}
\captionof{figure}{\label{wpivp}}
\end{center}
} \quad \hfill \parbox{60mm}{An example of a well-posed initial value problem is given in Figure \ref{wpivp}: from given values
$a,b,c,d$ one can calculate $e,f,g$ successively, they are uniquely determined. Indeed, the three dotted lines
are characteristics and they each intersect the initial value path in exactly one edge.}

\smallskip
\noindent
In both Figures \ref{quad1} and \ref{quad2} a finite piece of a quad-graph is given.
We assume these quad-graphs extend periodically in both the vertical and horizontal directions.
For equations on such a doubly infinite quad-graph there exist a two-parameter family of periodic reductions,
as for equations on the regular $\Z^2$-lattice. In fact, these quad-graphs are perturbations of the regular
$\Z^2$-lattice, we call them {\em irregular $\Z^2$-lattices}.

Theorem 1 extends to the more general setting of irregular $\Z^2$-lattices.
The proof is similar as in the $\Z^2$ setting: the fact the transfer matrix $\cL_{a,b}$
does not depend on the path from $a$ to $b$ follows from the Lax-condition.

\noindent
\parbox{75mm}{
\begin{center}
\setlength{\unitlength}{10mm}
\begin{picture}(6,4)(0,0)
\matrixput(0,0)(1,0){7}(0,1){5}{\circle{.12}}
\multiput(0,2)(3,0){3}{\circle*{.12}}
\multiput(1,2)(3,0){2}{\circle*{.12}}
\multiput(1,1)(3,0){2}{\circle*{.12}}
\multiput(2,1)(3,0){2}{\circle*{.12}}
\path(0,0)(2,2)(2,4)(0,4)
\path(1,0)(1,2)(0,2)(1,3)(1,4)(0,3)(0,1)(2,1)(2,0)(3,0)(5,2)(6,2)(6,1)(5,0)(5,1)(3,1)(2,0)(1,0)
\path(1,3)(3,3)(3,1)
\path(1,2)(3,4)(5,4)(5,2)
\path(2,2)(3,2)(4,3)(6,3)(6,2)
\path(3,3)(4,4)(4,3)
\path(4,2)(6,4)
\path(4,1)(4,0)(6,0)
\dottedline{.14}(1,.5)(2,.5)(3,1.5)(4,1.5)(5,2.5)(6,2.5)
\dottedline{.14}(1.5,0)(1.5,1)(2.5,2)(2.5,3)(3.5,4)
\dottedline{.14}(.5,3.5)(.5,2.5)(1.5,2.5)(1.5,1.5)(2.5,1.5)(2.5,.5)(3.5,.5)
\multiput(-.2,2.1)(3,0){3}{$b$}
\multiput(.8,2.1)(3,0){2}{$c$}
\multiput(.75,1.1)(3,0){2}{$d$}
\multiput(1.8,1.1)(3,0){2}{$a$}
\multiput(1.8,2.1)(3,0){2}{$e$}
\multiput(1.8,3.1)(3,0){2}{$f$}
\multiput(2.8,3.1)(3,0){2}{$g$}
\put(1.4,-.2){$p$}
\put(.8,.4){$q$}
\put(.4,3.5){$r$}
\Thicklines
\path(0,2)(1,2)(1,1)(2,1)(3,2)(4,2)(4,1)(5,1)(6,2)
\end{picture}
\captionof{figure}{\label{quad1}}
% Example of a well-posed periodic initial value problem on an irregular  $\Z^2$-lattice.}
\end{center}
} \quad \hfill \parbox{40mm}{In Figure \ref{quad1} the lattice parameters are attached to the edges as follows:
$p$ to the horizontal edges, $q$ to the vertical edges, and $r$ to the diagonal edges. From initial values $a,b,c,d$
the values $e,f,g$ are determined by
\begin{eqnarray*}
Q_{pr}(d,a,e,b)&=&0,\\
Q_{qr}(c,d,f,e)&=&0,\\
Q_{pq}(e,b,f,g)&=&0.
\end{eqnarray*}}

\smallskip
\noindent
We update values by shifting over $(2,1)$.
Thus, the mapping is
\[
(a,b,c,d)\mapsto (c,f,g,b),
\]
where
\[
f=d\frac{ar(rb-pd)+qc(rd-pb)}{aq(+rb-pd)+rc(rd-pb)},\quad
g=a\frac{rb(qa-pc)+qd(qc-ap)}{qb(qa-pc)+rd(qc-ap)}.
\]
A Lax pair $L,M$ for equation (\ref{H3}) is given by $L=L_{a,b}(p)$, $M=L_{a,c}(q)$, where
\[
L_{a,b}(p)=\frac{1}{a}\left(\begin{array}{cc} ka & -pab \\ p & -kb \end{array} \right),
\]
see \cite{WH}.
The monodromy matrix $L_{d,a}(p)L_{c,d}(q)L_{b,c}(p)L_{a,b}(r)$ yields one integral.

We perform the following reduction: in variables $x=a/c,y=b/d$ the mapping is expressed
\[
(x,y)\mapsto \left( \frac{p(qy+rx)-q(qxy+r)}{p(qx+ry)-q(rxy+q)}\frac{1}{x} ,
\frac{p(rx+qy)-r(q+rxy)}{p(qx+ry)-r(qxy+r)}\frac{1}{y} \right),
\]
and its integral is
\[
p^2 \left( \frac{x}{y}+\frac{y}{x} \right) - p(r+q)\left(x+y+\frac{1}{x}+\frac{1}{y} \right)
+ \left( xy + \frac{1}{xy} \right)qr.
\]

In the next example we need four lattice parameters to ensure that
the two lattice parameters on each quadrilateral differ.

\noindent
\parbox{75mm}{
\begin{center}
\setlength{\unitlength}{8mm}
\begin{picture}(8,5)(0,0)
\matrixput(0,0)(4,0){3}(0,4){2}{\circle{.15}}
\matrixput(2,1)(4,0){2}(0,4){2}{\circle{.15}}
\multiput(1,2)(1,0){3}{\circle{.15}}
\multiput(5,2)(1,0){3}{\circle{.15}}
\multiput(2,3)(4,0){2}{\circle{.15}}
\multiput(0,0)(4,0){2}{\circle*{.15}}
\multiput(1,2)(1,-1){2}{\circle*{.15}}
\put(0,4){\circle*{.15}}
\path(2,1)(2,3)(4,4)(2,5)(0,4)(2,3)
\path(1,2)(3,2)(4,4)(5,2)(4,0)(3,2)
\path(4,0)(5,2)(4,4)(6,5)(8,4)(7,2)(8,0)(6,1)(4,0)
\path(6,1)(6,3)(4,4)
\path(6,3)(8,4)
\path(5,2)(7,2)
\dottedline{.14}(1,.5)(1.5,2)(1,3.5)(3,4.5)
\dottedline{.14}(3,.5)(2.5,2)(3,3.5)(1,4.5)
\dottedline{.14}(5,.5)(5.5,2)(5,3.5)(7,4.5)
\dottedline{.14}(7,.5)(6.5,2)(7,3.5)(5,4.5)
\dottedline{.14}(.5,1)(2,1.5)(3.5,1)(4.5,3)(6,2.5)(7.5,3)
\dottedline{.14}(.5,3)(2,2.5)(3.5,3)(4.5,1)(6,1.5)(7.5,1)
\put(-.3,4.1){$a$}
\put(1,2.2){$b$}
\put(.2,-.2){$c$}
\put(2.2,1){$d$}
\put(3.6,-.2){$a$}
\put(2.1,2.1){$e$}
\put(2,3.3){$f$}
\put(3.2,2){$g$}
\put(4,4.2){$h$}
\put(5,2.2){$i$}
\put(8,0.2){$h$}
\put(6.2,1){$j$}
\put(2.2,5){$j$}
\put(3.1,4.5){$p$}
\put(.7,4.5){$q$}
\put(7.1,4.5){$q$}
\put(4.7,4.5){$p$}
\put(.2,2.9){$s$}
\put(.2,.9){$r$}
\Thicklines
\path(0,4)(1,2)(0,0)(2,1)(4,0)
\end{picture}
\captionof{figure}{\label{quad2} }
%Example of a well-posed periodic initial value problem on an irregular $\Z^2$-lattice.}
\end{center}
} \quad \hfill \parbox{40mm}{Given initial values $a,b,c,d$,
the values $e,f,g,h,i$ can be determined by the following equations:
\begin{eqnarray*}
Q_{rp}(b,c,e,d)&=&0,\\
Q_{sp}(a,b,f,e)&=&0,\\
Q_{rq}(e,d,g,a)&=&0,\\
Q_{sq}(f,e,h,g)&=&0,\\
Q_{rs}(g,a,h,i)&=&0,\\
Q_{qp}(a,f,j,h)&=&0.
\end{eqnarray*}
}

\smallskip
\noindent
The values $a,b,c,d$ are repeated periodically on the 'staircase' which extends
along a 'diagonal' of the quad-graph.The particular way of choosing lattice
parameters ($q,p,p,q$, see Figure \ref{quad2}, and we also take $r,s,s,r$ vertically)
ensures that the periodic solution has the same period as the initial values.
It is important to notice that when going one step to the right (on the $\Z^2$ part
of the quad-graph) the lattice parameters $p,r$ are interchanged with $q,s$ respectively.
Therefore, we consider the mapping
\[
(a,b,c,d,p,q,r,s)\mapsto(h,i,a,j,q,p,s,r).
\]
After the transformation $x=a/c,y=b/d$ we are left with
\[
(x,y,p,q,r,s)\mapsto(\frac{1}{x}\frac{x_n}{x_d},\frac{1}{y}\frac{y_n}{y_d},q,p,s,r),
\]
with
\begin{eqnarray*}
x_n&=&-qsp^2+pq(qs+pr)x+2ypsrq-(rq+ps)(qs+pr)yx-y^2sr^2q \\
&&-pq^2x^2r+sr(qs+pr)y^2x+(p^2q^2+s^2r^2)yx^2-y^2ps^2x^2r, \\
x_d&=&y^2rqp^2x-y^2rq^2p-y^2qsx^2p^2+y^2q^2sxp+yr^2s^2-ypqr^2x+2yrqsx^2p \\
&&-yp^2sxr-yq^2sxr+yq^2p^2-yps^2xq-r^2qsx^2+r^2spx+rs^2xq-ps^2r, \\
y_n&=&(s^2p-s(qs+pr)x-s^2ry+(prq+s^3)yx+x^2rqs-yqsx^2p) \\
&&\cdot(-qsp+q(qs+pr)x+ysrq-(srp+q^3)yx-q^2x^2r+x^2q^2py),\\
y_d&=&(-qsp+(prq+s^3)x+ysrq-s(qs+pr)yx-x^2s^2r+x^2s^2py)\\
&&\cdot(q^2p-(srp+q^3)x-yq^2r+q(qs+pr)yx+x^2rqs-yqsx^2p).
\end{eqnarray*}
The monodromy matrix,
\[
L_{d,a}(q)L_{c,d}(p)L_{b,c}(r)L_{a,b}(s)
\]
yields one invariant, in reduced variables,
\[
x(1+y^2)(qs+pr)+y(x^2+1)(pq+sr)-(x^2+y^2)rq-(x^2y^2+1)ps.
\]
The linear part of the mapping is easily solved by
\[
p=c_1+c_2(-1)^n,\ q=c_1-c_2(-1)^n,\ r=c_3+c_4(-1)^n,\ s=c_3-c_4(-1)^n
\]
where $n$ is an integer counting the iterations of the map.
Thus we may obtain an alternating two-dimensional map with one integral, cf. \cite{Q}.
However, this map would take too much space to write explicitly.
For special values of the parameters and one of the initial values
we have observed that the growth of this map is quadratic, which indicates
it is integrable.

\section{Multivaluedness of iterates of correspondences} \label{mc}
We have seen two examples in which correspondences arose: ($n,0$)-reductions of the QD-system in section
\ref{qdh} and of the pKdV-equation in section \ref{pkdvh}.
In general the multi-valuedness of the iterates of a correspondence would grow exponentially.
However, for integrable correspondences one expects the multi-valuedness to grow polynomially
instead.

\subsection{The pKdV (3,0)-correspondence}
\noindent
\parbox{55mm}{
\setlength{\unitlength}{6mm}
\begin{center}
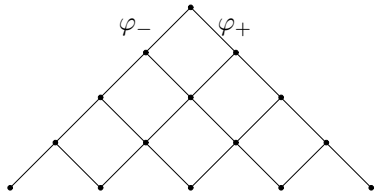

\begin{picture}(8,4)(0,0)
\multiput(0,0)(2,0){5}{\circle*{.1}}
\multiput(1,1)(2,0){4}{\circle*{.1}}
\multiput(2,2)(2,0){3}{\circle*{.1}}
\multiput(3,3)(2,0){2}{\circle*{.1}}
\put(4,4){\circle*{.1}}
\path(0,0)(4,4)(8,0)
\path(1,1)(2,0)(5,3)
\path(2,2)(4,0)(6,2)
\path(3,3)(6,0)(7,1)
\put(2.4,3.5){$\mapt_-$}
\put(4.6,3.5){$\mapt_+$}
\end{picture}
\captionof{figure}{\label{blc} Polynomial growth of multivaluedness of iterates.}
\end{center}
} \quad \hfill \parbox{60mm}{
Generically the multi-valuedness of the $l$th iterate of a two-valued
map would be $2^l$. This is not the case here. Due to the $x\leftrightarrow
y$ symmetry of the system (\ref{ste}) the correspondence $\mapt$ equals its own
inverse. The relations $\mapt^{-1}_{\pm}=\mapt_{\mp}$ imply that the $l$th iterate of
the correspondence is ($l+1$)-valued, see Figure \ref{blc}. All points on the same
vertical line have the same value.}

\subsection{The QD ($n,0$)-correspondence}
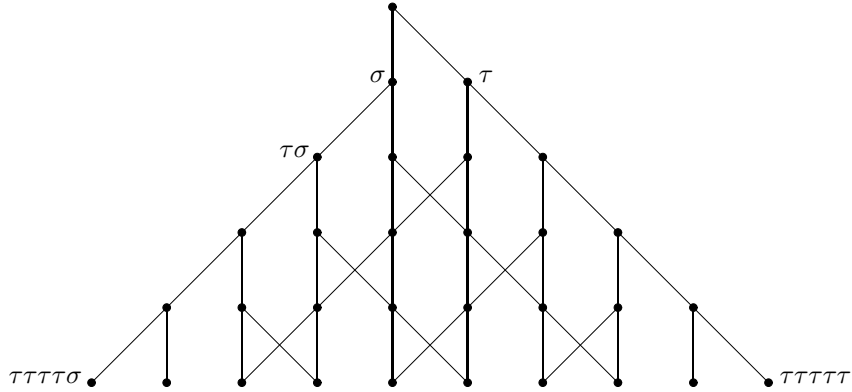
\begin{figure}[h]
\setlength{\unitlength}{1cm}
\begin{center}
\begin{picture}(9,5)(0,0)
\multiput(0,0)(1,0){10}{\circle*{.1}}
\multiput(1,1)(1,0){8}{\circle*{.1}}
\multiput(2,2)(1,0){6}{\circle*{.1}}
\multiput(3,3)(1,0){4}{\circle*{.1}}
\multiput(4,4)(1,0){2}{\circle*{.1}}
\put(4,5){\circle*{.1}}
\multiput(0,0)(1,1){4}{\line(1,1){1}}
\multiput(2,0)(1,1){3}{\line(1,1){1}}
\multiput(4,0)(1,1){2}{\line(1,1){1}}
\multiput(6,0)(1,1){1}{\line(1,1){1}}
\multiput(1,0)(1,0){8}{\line(0,1){1}}
\multiput(2,1)(1,0){6}{\line(0,1){1}}
\multiput(3,2)(1,0){4}{\line(0,1){1}}
\multiput(4,3)(1,0){2}{\line(0,1){1}}
\put(4,4){\line(0,1){1}}
\put(3,0){\line(-1,1){1}}
\multiput(5,0)(-1,1){2}{\line(-1,1){1}}
\multiput(7,0)(-1,1){3}{\line(-1,1){1}}
\multiput(9,0)(-1,1){5}{\line(-1,1){1}}
\put(3.7,4){$\Pm$}
\put(5.15,4){$\Rm$}
\put(2.5,3){$\Rm\Pm$}
\put(-1.1,0){$\Rm\Rm\Rm\Rm\Pm$}
\put(9.15,0){$\Rm\Rm\Rm\Rm\Rm$}
\end{picture}
\caption{\label{2lf} Graphical representation of Proposition \ref{2l}}
\end{center}
\end{figure}

As one can easily verify the mappings $\Rm,\Pm$, given by equations (\ref{3pm},\ref{3rm}),
satisfy
\begin{equation} \label{PPP}
\Rm\Pm\Rm=\Pm\Pm\Pm\quad \text{and} \quad \Rm\Pm\Pm=\Pm\Pm\Rm.
\end{equation}
Due to these relations the $l$th iterate of the correspondence is $2l$ valued, see
figure \ref{2lf}.

\begin{proposition} \label{2l}
With compositions of mappings (\ref{PPP}), the $l$th iterate of the correspondence
($\sigma,\tau$) is $2l$ valued.
\end{proposition}
\proof The relations (\ref{PPP}) can be used to rewrite every word in two symbols
$\Pm$ and $\Rm$ of length $l$ into either a word that does not contain $\Rm\Pm$ or
a word that only contains $\Rm\Pm$ at the end. There are $2l$ such words:
\begin{align*}
&\Rm\Rm\Rm\cdots \Rm\Rm\Rm, \Pm\Rm\Rm\cdots \Rm\Rm\Rm, \Pm\Pm\Rm\cdots \Rm\Rm\Rm, \ldots , \Pm\Pm\Pm \cdots \Pm\Pm\Rm, \\
&\Rm\Rm\Rm\cdots \Rm\Rm\Pm, \Pm\Rm\Rm \cdots \Rm\Rm\Pm, \Pm\Pm\Rm\cdots \Rm\Rm\Pm, \ldots , \Pm\Pm\Pm\cdots \Pm\Pm\Pm.
\end{align*}
So there are at most $2l$ inequivalent words of length $l$. To show that there
are exactly $2l$ inequivalent words of length $l$ we proceed by induction.
We suppose there are are exactly $2l$ inequivalent words of length $l$. For
any three words $u,v,w$ we have $uv=uw\Rightarrow v=w$. So if two different words
of length $l$ extend to an equivalent word of length $l+1$, this word can be written (we
concatenate from the left) as $u = \tau v =\sigma w$. If a third word of length $l$ would
extend to $u$ then we have $u = \tau z$ or $u=\sigma z$ which would imply $z=v$ or $z=w$
respectively. Therefore words of length $l$ that extend to words of length $l+1$ coincide
at most pairwise, giving an lower bound of $2l$ on the number of words of length $l+1$.
However, since there are two words of length $l+1$ that are not equivalent to any other word,
namely $\Rm\Rm\Rm\cdots \Rm\Rm\Rm$ and $\Pm\Rm\Rm\cdots \Rm\Rm\Rm$, there are at least $2l+2$
inequivalent words of length $l+1$.
\qed

\noindent
The same relations hold in the $2n$-dimensional case.
\begin{proposition}
The $2n$-dimensional mappings $\Rm=\Rm_{2n}$ (\ref{Rm}) and $\Pm=\Pm_{2n}$ (\ref{Pm}) satisfy the
relations (\ref{PPP}).
\end{proposition}
\proof
That the second of the relations (\ref{PPP}) holds is easily established as
$\Pm\Pm$, i.e. the mapping $x_k\mapsto x_{k+2}$, clearly commutes with $\Rm$.

For the first relation we solve the equations (\ref{deqs}) for the $x_i$ in terms
of the $\h{x_k}$ to find the inverse $\Rm^{-1}$, with $i=1,2,\ldots,n$,
\begin{align*}
x_{2i-1} &\mapsto x_{2i-1} \frac{W^n_{2i}}{W^n_{2i-2}}, \\
x_{2i} &\mapsto x_{2i-2} \frac{W^n_{2i-4}}{W^n_{2i-2}},
\end{align*}
where
\begin{equation} \label{W}
W^n_k=\sum_{i=1}^{n-1} \left( \prod_{j=1}^{i-1} x_{2j-1+k} \right)
 \left( \prod_{j=i}^{n-1} x_{2j+2+k} \right) .
\end{equation}
One can verify that $\Pm(W^n_k)=Q^n_k$. This implies that $\Pm\Rm^{-1}$ is given by
\begin{align*}
x_{2i-1} &\mapsto x_{2i+2} \frac{Q^n_{2i}}{Q^n_{2i-2}} \\
x_{2i} &\mapsto x_{2i-3} \frac{Q^n_{2i-4}}{Q^n_{2i-2}}, \qquad i=1,\ldots,n
\end{align*}
which equals its inverse $\Rm\Pm^{-1}$. Multiplying $\Rm\Pm^{-1}=\Pm\Rm^{-1}$ from the left
by $\Rm\Pm$ and from the right by $\Pm$, using $\Rm\Pm\Pm=\Pm\Pm\Rm$, gives us
$\Rm\Pm\Rm=\Pm\Pm\Pm$. \qed

{\bf Remark:} One can also directly prove that the $\Rm^{-1}$ provided is the inverse of $\Rm$.
This relies on the fact that $\Rm(W^n_k)=Q^n_k$, which in terms of $Q$ amounts to the identity
\[
\prod_{i=1}^{n-1} Q^n_{2i} = \sum_{i=1}^n (\prod_{j=1}^{i-1} x_{2j-1})(
\prod_{j=i}^{n-1} x_{2j+4})( \prod_{j=1}^{n-2} Q^n_{2j+2i}),
\]
which can be proved using
\[
Q^{n-m}_0Q^n_2= x_1Q^{n-m-1}_2Q^n_0 + Q^n_{2(n-m)} (\prod_{i=3}^{n-m+1} x_{2i})
\]
which in turn relies on a generalization of the relations (\ref{rq}),
\[
Q^{n-m}_0=Q^{n-m}_{2m}(\prod_{i=2}^{m+1} x_{2i})
+ (Q^{m}_{0}\prod_{i=m}^{n-1} x_{2i+1}).
\]

\section{Concluding remarks}
We have shown that the staircase method provides integrals for
mappings and correspondences derived as $\s$-periodic reductions
of lattice equations and systems of lattice equations.
We also showed that such mappings and correspondences
can be order-reduced systematically, using
symmetries of the lattice equations.
In all examples encountered the staircase method yields
sufficiently many functionally independent integrals for the
$d$-reduced mappings and correspondences to be completely integrable.
However, we know the above statement is not true in general.
In \cite{SNK} periodic reductions of systems of P$\Delta$Es
are considered for which the staircase method does not provide
sufficiently many integrals. However, in those cases it was observed that
a certain linear combination of integrals factorises into a product of
$2$-integrals, from which then another integral can be obtained.

For posing initial value problems we used the method laid out
in \cite{K}. Although, for the Boussinesq system we presented
an alternative approach. We have calculated integrals for
mappings/correspondences with dimension up to say 20.
Therefore we were able to check functional independence of the
integrals using the symbolic software package Maple \cite{Map}.

Closed form expressions for integrals of mappings with arbitrary
dimensions (corresponding to reductions with $s_2=-1$) have been
given in \cite{KRQ,TKQ}. There, the lattice
equations are of Adler-Bobenko-Suris type \cite{ABS1,ABS2} and
the integrals are expressed in terms of multisums of products.
Their functional independence and involutivity is being investigated \cite{Inv}.

It is an open problem whether such closed form expressions can be
obtained in general. In particular, we don't know whether such
expressions can be given for the Bruschi-Calogero-Droghei equation,
the Quotient-Difference algorithm and the Boussinesq system presented
here, except for QD in the case $s_1=0$. Another question is how to obtain
symplectic structures for the mappings studied in this paper. This would
enable one to conclude complete integrability.

We have obtained a few results on mappings of arbitrary dimension. These
include an explicit $(2+(-1)^{n})$-reduction of the mapping related to
the $(n-1,1)$-reduction of the lattice potential KdV equation, as well
as an explicit expression for a $2$-integral of the mapping in the case
that $n$ is even. We proved that the
$(n-1,1)$-reduction of the Boussinesq system can be $d$-reduced with $d=6$
if 3 divides $n$ and with $d=2$ otherwise. Also, we presented two $3$-integrals for this
mapping (expressed in terms of the original variables). For the $(n-1,-1)$-reduction
of the Boussinesq system we have showed that the mapping can be $d$-reduced
with $d=6$ if 3 divides $n+1$ and with $d=2$ otherwise. Finally, the $(n,0)$-reduction
of the QD-system yields a $2n$-dimensional $2$-valued correspondence. We have given
an explicit expression for this correspondence and showed that its $n$th iterate
is $2n$-valued. It would be interesting to investigate other ways of establishing
such a result because one does not always have explicit expressions at hand.

\vspace{5mm}
\noindent
{\bf Acknowledgement.}

\noindent
This research has been funded by the Australian Research Council
through the Centre of Excellence for Mathematics and Statistics of
Complex Systems. We are grateful for the hospitality of the Isaac Newton
Institute during the 2009 program on Discrete Integrable Systems.


\begin{thebibliography}{10}

\bibitem{AL}
M.J. Ablowitz and F.J. Ladik,
A nonlinear difference scheme and inverse scattering, Stud. Appl.
Math. 55 (1976) 213--229;
On the solution of a class of nonlinear partial difference equations,
ibid. 57 (1977) 1-12.

\bibitem{ABS1}
V.E. Adler, A.I. Bobenko, Yu.B. Suris,
Classification of integrable equations on quad-graphs. The consistency approach,
Commun. Math. Phys. 233 (2003), 513--543. arXiv:nlin/0202024.

\bibitem{ABS2}
V.E. Adler, A.I. Bobenko, Yu.B. Suris,
Discrete nonlinear hyperbolic equations. Classification of integrable cases,
Funct. Anal. Appl. 43 (2009), 3--21. arXiv:nlin/0705.1663

\bibitem{AV}
V.E. Adler and A.P. Veselov,
Cauchy Problem for Integrable Discrete Equations on Quad-Graphs
Acta Applicandae Mathematicae 84: 237--262, 2004.

\bibitem{Brez}
C. Brezinski,
Pad\'{e}-type approximation and general orthogonal polynomials,
International Series in Numerical Mathematics, Birkhauser, Basel, 1980.

\bibitem{BCD}
M. Bruschi, F. Calogero and R. Droghei,
Tridiagonal matrices, orthogonal polynomials and Diophantine relations I,
J. Phys. A: Math. Theor. 40 (2007), 9793--9817.

\bibitem{BRSZ}
M. Bruschi, O. Ragnisco, P. M. Santini, and T.G. Zhang,
Integrable symplectic maps,
Physica D 49 (1991), 273--294.

\bibitem{Cabe}
J.H. McCabe,
The quotient difference algorithm and the Pad\'{e} table: An alternative form and a general continued fraction,
Mathematics of Computation 41(6) (1983), 183--197.

\bibitem{CN}
F. Calogero and M. C. Nucci, Lax pairs galore, J. Math. Phys. 32 (1991) 72--74.

\bibitem{CNP}
H.W. Capel, F.W. Nijhoff and V.G. Papageorgiou,
Complete Integrability of Lagrangian Mappings and Lattices of KdV Type,
Physics Letters 155A (1991), 377--387.

\bibitem{DJM} E. Date, M. Jimbo and T. Miwa,
Method for Generating Discrete Soliton Equations I-V, J. Phys.
Soc. Japan 51 (1982), 4116--4124, 4125--4131, 52 (1983), 388--393, 761--765, 766--771.

\bibitem{BF}
 B. Fuchssteiner, Coupling of completely integrable system: the perturbation bundle.
 In: P.A. Clarkson, Editor, Applications of analytic and geometric methods to nonlinear
 differential equations, Kluwer, Dordrecht (1993) 125--138.

\bibitem{HBQC}
F. Haggar, G.B. Byrnes, G.R.W. Quispel and H.W. Capel,
$k$-integrals and $k$-Lie symmetries in discrete dynamical systems,
Physica 233A (1996) 379--394.

\bibitem{MH}
M. Hay, Discrete Lax pairs, Reductions and Hierarchies,
PhD thesis, The University of Sydney (2008).

\bibitem{HW}
P. Henrici and Bruce O. Watkins,
Finding zeros of a polynomial by the Q-D algorithm,
Communications of the ACM 8(9) (1965), 570--574.

\bibitem{WH}
W. Hereman, Symbolic computation of Lax Pairs of integrable nonlinear
difference equations on quad-graphs,
http://www.newton.ac.uk/programmes/DIS/seminars/062914001.pdf.

\bibitem{H} R. Hirota,
Nonlinear Partial Difference Equations I-III,
J. Phys. Soc. Japan 43 (1977) 1424--1433, 2074--2078, 2079--2089.

\bibitem{Map}
Maple, http://www.maplesoft.com/products/Maple/index.aspx

\bibitem{NTS}
A. Nagai, T. Tokihiro and J. Satsuma,
The Toda Molecule Equation and the $\epsilon$-Algorithm,
Mathematics of Computation 67 (1998) 1565--1575

\bibitem{NC}
F.W. Nijhoff and H.W. Capel,
The Discrete Korteweg-de Vries equation,
Acta Applicandae Mathematicae 39 (1995) 133--158.

\bibitem{NPC}
F.W. Nijhoff, V.G. Papageorgiou and H.W. Capel,
Integrable Time-Discrete Systems: Lattices and Mappings,
Ed. P.P. Kulish, in: Quantum Groups, Lecture Notes in Mathematics vol. 1510, pp.
312--325, Berlin/New York, Springer Verlag, 1992.

\bibitem{NPCQ}
F.W. Nijhoff, V.G. Papageorgiou, H.W. Capel and G.R.W. Quispel,
The Lattice Gel'fand-Dikii Hierarchy,
Inverse Problems 8 (1992) 597--621.

\bibitem{NQC} F.W. Nijhoff, G.R.W. Quispel and H.W. Capel,
Direct linearization of nonlinear difference-difference equations,
Phys. Lett., 97A (1983) 125--128.

\bibitem{NW}
F.W. Nijhoff and A.J. Walker,
The discrete and continuous Painlev\'e VI hierarchy and the Garnier systems,
Glasgow Math. J. 43A (2001), 109--123.

\bibitem{PGR}
V. Papageorgiou, B. Grammaticos and A. Ramani,
Orthogonal polynomial approach to discrete Lax pairs for initial boundary-value problems
of the QD algorithm,
Letters in Mathematical Physics 34(2) (1995), 91--101.

\bibitem{PN}
V.G. Papageorgiou and F.W. Nijhoff, On some Integrable Discrete-Time
Systems associated with the Bogoyavlensky Lattices,
In: Statistical Mechanics, Soliton Theory and Nonlinear Dynamics, Festschrift in
honour of H.W. Capel's 60th Birthday, eds. G.R.W. Quispel, F.W.
Nijhoff and J.H.H. Perk, Physica 228A (1996), 172--188.

\bibitem{PNC}
V.G. Papageorgiou, F.W. Nijhoff, and H.W. Capel,
Integrable mappings and nonlinear integrable lattice equations,
Phys. Lett. A 147 (1990) 106--114.

\bibitem{Q}
G.R.W. Quispel,
An alternating integrable map whose square is the QRT map,
Phys. Lett. A 307 (2003) 50–4.

\bibitem{QCPN}
G.R.W. Quispel, H.W. Capel, V.G. Papageorgiou and F.W. Nijhoff,
Integrable mappings derived from soliton equations,
Physica A 173 (1991) 243--266.

\bibitem{QNCL} G.R.W. Quispel, F.W. Nijhoff, H.W. Capel and J. van der Linden,
Linear Integral Equations and Nonlinear Difference-Difference Equations,
Physica 125A (1984) 344--380.

\bibitem{QRT1}
G.R.W. Quispel, J.A.G. Roberts and C.J. Thompson,
Integrable Mappings and Soliton Equations I, Phys. Lett. 126A (1988) 419-421.

\bibitem{QRT2}
G.R.W. Quispel, J.A.G. Roberts and C.J. Thompson,
Integrable Mappings and Soliton Equations II, Physica 34D (1989) 183-192.

\bibitem{RKQ}
O. Rojas, P.H. van der Kamp and G.R.W. Quispel,
Lax representations for integrable mappings,
In preparation.

\bibitem{SNK}
P.E. Spicer, F.W. Nijhoff, P.H. van der Kamp,
Higher analogues of the
discrete-time Toda equation and the quotient-difference algorithm (2010) arXiv:nlin/1005.0482.

\bibitem{TN}
A.S. Tongas and F.W. Nijhoff,
The Boussinesq integrable system. Compatible lattice and continuum structures,
Glasgow Math. J. 47A (2005) 205--219. arXiv:nlin/0402053

\bibitem{TKQ}
D. Tran, P.H. van der Kamp, G.R.W. Quispel,
Closed-form expressions for integrals of traveling wave reductions of integrable lattice equations,
J. Phys. A: Math. Theor. 42 (2009) 225201.

\bibitem{Inv}
D. Tran, P.H. van der Kamp, G.R.W. Quispel, Involutivity of sine-Gordon, modified and potential KdV maps (2010) arXiv:nonl/1010.3471.

\bibitem{K}
P.H. van der Kamp,
Initial value problems for lattice equations, J. Phys. A: Math. Theor. 42 (2009) 404019.

\bibitem{grods}
P.H. van der Kamp,
Growth of degrees of integrable mappings, Journal of Difference Equations and Applications (2010)
arXiv:nonl/1005.2065v1.

\bibitem{KRQ}
P.H. van der Kamp, O. Rojas and G.R.W. Quispel,
Closed-form expressions for integrals of mKdV and sine-Gordon maps,
J. Phys A: Math Gen. 40 (2007) 12789—12798.

\bibitem{Ves1}
A.P. Veselov,
Growth of the number of images of a point under iterates of a multivalued map,
Math. Notes 49 (1991), 134--139.

\bibitem{Ves}
A.P. Veselov,
Integrable maps,
Russian Mathematical Surveys 46 (1991), 1--51.



\end{thebibliography}
\end{document}